# Synthetic control over the binding configuration of luminescent *sp*³-defects in single-walled carbon nanotubes

Simon Settele[1], Felix J. Berger[1,2], Sebastian Lindenthal[1], Shen Zhao[3], Abdurrahman Ali El Yumin[1,2], Nicolas F. Zorn[1,2], Andika Asyuda[1], Michael Zharnikov[1], Alexander Högele[3,4] & Jana Zaumseil[1,2 ✉]

The controlled functionalization of single-walled carbon nanotubes with luminescent *sp*³-defects has created the potential to employ them as quantum-light sources in the near-infrared. For that, it is crucial to control their spectral diversity. The emission wavelength is determined by the binding configuration of the defects rather than the molecular structure of the attached groups. However, current functionalization methods produce a variety of binding configurations and thus emission wavelengths. We introduce a simple reaction protocol for the creation of only one type of luminescent defect in polymer-sorted (6,5) nanotubes, which is more red-shifted and exhibits longer photoluminescence lifetimes than the commonly obtained binding configurations. We demonstrate single-photon emission at room temperature and expand this functionalization to other polymer-wrapped nanotubes with emission further in the near-infrared. As the selectivity of the reaction with various aniline derivatives depends on the presence of an organic base we propose nucleophilic addition as the reaction mechanism.

[1] Institute for Physical Chemistry, Universität Heidelberg, Heidelberg, Germany. [2] Centre for Advanced Materials, Universität Heidelberg, Heidelberg, Germany. [3] Fakultät für Physik, Munich Quantum Center and Center for NanoScience (CeNS), Ludwig-Maximilians-Universität München, München, Germany. [4] Munich Center for Quantum Science and Technology (MCQST), München, Germany. ✉email: zaumseil@uni-heidelberg.de

Covalent functionalization of single-walled carbon nanotubes (SWNTs) has a long history and many different types of chemistry have been explored[1–3]. However, a major paradigm shift occurred with the discovery of luminescent $sp^3$ defects[4–10]. Rather than quenching near-infrared photoluminescence (PL) of semiconducting SWNTs, these $sp^3$ defects with covalently attached functional groups were found to create highly emissive states at lower energies. Mobile excitons, which result in $E_{11}$ emission if not quenched[11], are funneled to these defect states with long PL lifetimes and thus the usually low PL quantum yield (PLQY) of SWNTs increases significantly[4,12,13]. Luminescent defects (or organic color centers)[7] have expanded the emission range of SWNTs further toward the near-infrared and opened up applications in sensing[14,15], in vivo imaging within the second biological window[16,17], and as quantum-light sources with room-temperature single-photon emission[18–22]. Various synthetic methods (e.g., radical-based reactions with aryl diazonium salts) allow for different functional groups to be attached to the SWNTs, which lead to additional small energetic shifts[18,23–26]. However, defect emission in general occurs over a wide spectral range (e.g., from 1100 to 1350 nm for (6,5) SWNTs). Indeed, the emission wavelength of $sp^3$-functionalized SWNTs is less determined by the nature of the attached functional group than by the precise defect-binding configuration within the $sp^2$-hybridized SWNT lattice[20,24]. For any functional group that forms a bond with the nanotube, thus creating one $sp^3$ carbon, another carbon atom, either in ortho or para position to the first, must be saturated (e.g., with another substituent, -OH or -H group). For chiral nanotubes, this requirement leads to six possible binding configurations. Each of these is associated with a distinct defect energy and thus emission wavelength as corroborated by density functional theory calculations[9,27,28]. Most functionalization methods produce a mix of configurations and thus defect emission bands, the most common being the $E_{11}^\star$ and the even more red-shifted $E_{11}^{\star -}$ emission with a longer PL lifetime[29–31]. They are considered to originate from the ortho-L$_{90}$ (ortho++, circumferential direction) and ortho-L$_{30}$ (ortho+, orientation along the nanotube axis) configurations of the two $sp^3$ carbons, respectively[20,24,27]. Spectral variation is a problem for practical applications and must be reduced to ideally one single type of defect emission with narrow wavelength and PL lifetime distributions for the entire ensemble of functionalized nanotubes.

The more red-shifted $E_{11}^{\star -}$ emission is closer to telecommunication wavelengths and is better suited for high-purity single-photon emission at room temperature. However, it is usually found for only a small portion of functionalized nanotubes. Saha et al.[20] suggested the use of achiral zigzag-SWNTs whose symmetry leads to degenerate defect configurations and hence single peak emission. However, zigzag nanotubes are a very rare species in typical nanotube raw materials[32] and thus difficult to obtain and purify in significant amounts. Another approach is the functionalization of SWNTs with either divalent functional groups (e.g., >CF$_2$)[29] or bidentate reactants with short bridging moieties (e.g., bisdiazonium compounds)[31], which increase the probability of a specific binding configuration, albeit with limitations. Gaining real synthetic control over the defect-binding configuration and hence emission wavelength in a simple, reproducible, and possibly scalable manner is highly desirable.

Here we demonstrate a synthetic approach for the introduction of luminescent $sp^3$ defects in carbon nanotubes, with which we can either create both types of binding configurations leading to $E_{11}^\star$ and $E_{11}^{\star -}$ emission, or exclusively the configuration for $E_{11}^{\star -}$ emission at longer wavelengths and with longer PL lifetimes. We employ polymer-sorted monochiral (6,5) SWNTs for this functionalization[33], as they are easily purified and are available in relatively large amounts. The use of polymer-wrapped SWNTs in organic solvents avoids the limitations of reactions in aqueous dispersions[10] and, hence, dramatically expands the available chemical toolbox. The selectivity and reactivity of the functionalization step is controlled by the concentration of potassium *tert*-butoxide (KO$^t$Bu) as a base and, thus, we propose nucleophilic addition as the underlying reaction mechanism as opposed to the commonly applied radical-based functionalization routes that mainly produce defects for $E_{11}^\star$ emission.

## Results

**Synthetic control over defect emission from (6,5) SWNTs.** Polymer-wrapped (6,5) SWNTs (length 1–2 μm) serve as the model system for the controlled introduction of different luminescent $sp^3$ defects in this study and were selectively dispersed and purified by shear-force mixing in toluene with a polyfluorene copolymer (poly[(9,9-dioctylfluorenyl-2,7-diyl)-*alt*-(6,6')-(2,2'-bipyridine)] (PFO-BPy), see Fig. 1a and "Methods"). Excess polymer was removed by vacuum filtration and all reactions were performed with (6,5) nanotubes redispersed in fresh toluene (Supplementary Fig. 1). Functionalization occurred via base-mediated coupling with 2-haloanilines (2-iodoaniline if not specified otherwise) as shown in Fig. 1a, either under ultraviolet (UV)-light irradiation (~365 nm, blue reaction path) or in the dark (red reaction path). As we will see, the key component for controlling the reaction is the organic base KO$^t$Bu. Tetrahydrofuran (THF, 8.3 vol%) and dimethyl sulfoxide (DMSO, 8.3 vol%) were added as co-solvents to increase reactivity and selectivity. The functionalization was performed in an open flask at room temperature (10–180 min). Vacuum filtration of the mixture quenched the reaction by separating the reagents from the nanotubes. The collected SWNTs were washed with methanol and redispersed again in toluene for characterization. For a detailed step-by-step description, see Supplementary Methods 1 and Supplementary Table 1.

Nanotubes that were functionalized with 2-iodoaniline under UV illumination exhibited PL spectra (Fig. 1b) with the $E_{11}$ emission of the mobile excitons and two red-shifted emission bands at ~1130 nm and ~1250 nm, which we assign to $E_{11}^\star$ and $E_{11}^{\star -}$, respectively. The relative intensities of the two defect emission bands could be controlled by the amount of KO$^t$Bu (0.5–3 eq.) in the reaction mixture, whereas the concentration of 2-iodoaniline was kept constant. With increasing base concentration, the reactivity increased (i.e., higher ratio of defect to $E_{11}$ intensity) and the $E_{11}^{\star -}$ emission became more and more dominant. Evidently, a higher base concentration favors the creation of the defect-binding configuration corresponding to $E_{11}^{\star -}$. Nevertheless, the defect configuration for $E_{11}^\star$, which is the majority product for reactions with diazonium salts[34] as well as photoactivated reactions of 4-iodoaniline in aqueous dispersion[23], is still significant at lower base concentrations as long as UV illumination is applied.

Surprisingly, when the same reaction was performed in the dark (i.e., without UV-light activation), only the more red-shifted $E_{11}^{\star -}$ emission feature (peak width 65 meV) was observed (Fig. 1c). The emission intensity and, thus, the number of defects increased steadily with base concentration and reaction time, and could easily be tuned using these parameters (Fig. 1c, d). Even after very long reaction times, no other defect emission bands appeared. Concentration-dependent functionalization reactions revealed that the highest reactivity and selectivity for defects with $E_{11}^{\star -}$ emission is achieved when the aniline reagent is used in excess compared to the SWNTs (see Supplementary Note 1 and Supplementary Fig. 2). To the best of our knowledge, this reaction constitutes the first example of $sp^3$ functionalization of chiral carbon nanotubes that exhibits such a high selectivity for this specific type of defect-binding configuration and exclusively leads to strongly red-shifted emission ($E_{11}^{*-}$).

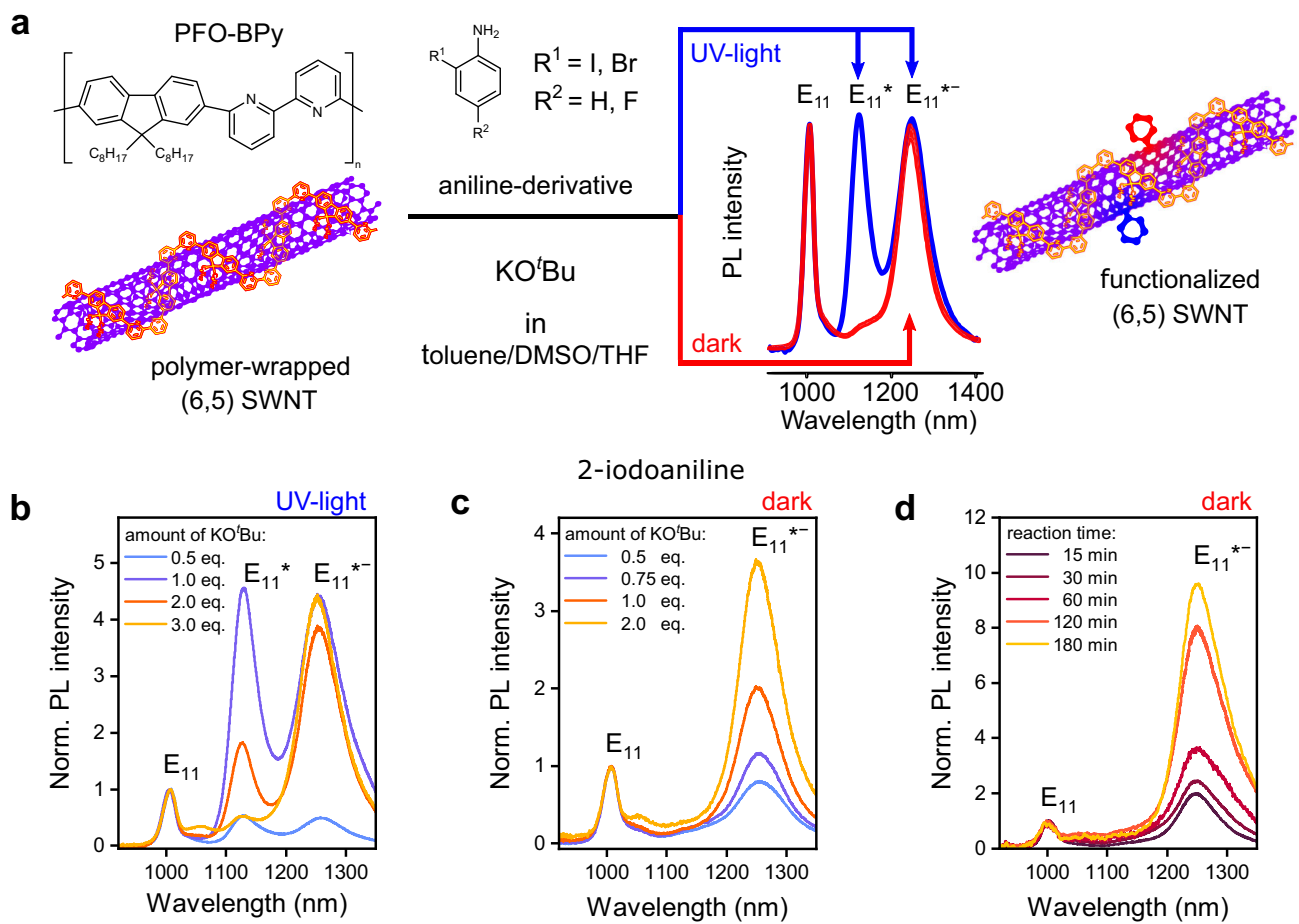

**Fig. 1 Controlling $sp^3$-defect emission from (6,5) SWNTs. a** Scheme for functionalization of PFO-BPy-wrapped (6,5) SWNTs in toluene/DMSO/THF with aniline derivatives and base KO$^t$Bu under UV-light irradiation (blue) and in the dark (red), respectively. Red-shifted defect emission bands are labeled as $E_{11}^*$ (~1130 nm) and $E_{11}^{*-}$ (~1250 nm). **b, c** PL spectra of (6,5) SWNTs functionalized with 2-iodoaniline and varying concentrations of KO$^t$Bu under UV-light irradiation (**b**) and in the dark (**c**); reaction time 30 min. **d** Evolution of $E_{11}^{*-}$ emission normalized to $E_{11}$ intensity for increasing reaction times with 2 eq. of KO$^t$Bu in the dark. The concentration of 2-iodoaniline was kept at 29.30 mmol L$^{-1}$.

The established reaction scheme was not limited to 2-iodoaniline but was also performed successfully with 2-bromoaniline and 5-fluoro-2-iodoaniline (see Supplementary Fig. 3), although with different reactivities. A high selectivity and yield for $E_{11}^{*-}$ defects were achieved with 2-bromoaniline, whereas the additional steric repulsion in 5-fluoro-2-iodoaniline slightly lowered reactivity and selectivity. Overall, the base-mediated functionalization of nanotubes is selective, scalable, and uses inexpensive reagents that are easy and safe to handle. By employing this procedure, we were able to obtain high-quality dispersions of functionalized (6,5) SWNTs with only $E_{11}^{*-}$ defects, which previously were only accessible in conjunction with $E_{11}^*$ defects. Before investigating the origin of this selectivity, the spectroscopic properties of these functionalized nanotubes and their possible application as single-photon emitters will be discussed and demonstrated.

**Spectroscopic properties of functionalized (6,5) SWNTs.** As shown above (Fig. 1d), the reaction time represents an excellent tool to precisely tune the $E_{11}^{*-}$ defect density of polymer-wrapped (6,5) SWNTs functionalized with 2-iodoaniline in the dark, which is reflected in the linear correlation of the $E_{11}^{*-}/E_{11}$ PL intensity ratio (Fig. 2a). Various additional metrics can be used to quantify the degree of functionalization. These are the integrated $E_{11}^{*-}/E_{11}$ absorption peak ratio and the Raman D/G$^+$-mode ratio. Both are expected to be directly proportional to the number of $sp^3$ defects[35]. The $E_{11}^{*-}$ defect absorption band, located around ~1247 nm (Fig. 2b), is much weaker than the $E_{11}$ absorption but increases steadily with reaction time. The integrated $E_{11}^{*-}/E_{11}$ absorption ratios for different reaction times also reveal a linear correlation with the $E_{11}^{*-}/E_{11}$ PL intensity ratios (Fig. 2c). Their large difference (approximately a factor of 100) indicates the strong funneling effect of mobile excitons toward the luminescent defect states. The Stokes shift between $E_{11}^{*-}$ absorption and emission was rather small with 9–16 meV, similar to values for $E_{11}^*$ defects introduced by diazonium chemistry[34]. The systematic increase of the Raman D-mode signal and thus D/G$^+$ mode ratio with reaction time also showed a linear correlation with the $E_{11}^{*-}/E_{11}$ emission ratio (Supplementary Fig. 4) and can be used as an independent metric for the defect density.

As previously shown for $E_{11}^*$ defects[4,34], the controlled functionalization of SWNTs can be used to significantly enhance their ensemble PLQYs, albeit only at optimum defect densities. PLQY data (absolute values obtained with an integration sphere, see "Methods") for (6,5) SWNTs with $E_{11}^{*-}$ defects are shown in Fig. 2d. The intensity of the $E_{11}$ emission and thus also its contribution to the total PLQY decreases steeply with the number of defects (here quantified by the Raman D/$G^+$ ratio), whereas the $E_{11}^{*-}$ share rises sharply as mobile excitons are trapped by the emissive defects. At even higher defect densities, the absolute $E_{11}$ and $E_{11}^{*-}$ emission and thus PLQY decreases again as the $sp^2$ lattice is disrupted by too many $sp^3$ carbons, preventing the formation of extended electronic states. Areas with such high $sp^3$ carbon concentrations act as quenching sites, thus decreasing the overall number of excitons for emission. Nevertheless, a strong brightening effect compared to the pristine samples (PLQY ~ 2%) was observed for the total PLQY of functionalized nanotubes, reaching up to 5.1% at optimal defect density. This represents a significant improvement compared to polymer-wrapped (6,5) SWNTs functionalized with diazonium salts (mainly $E_{11}^{*}$ defects) and is consistent with the larger optical trap depth (difference between $E_{11}$ and defect emission energy) of the $E_{11}^{*-}$ defects (241–247 meV)[34].

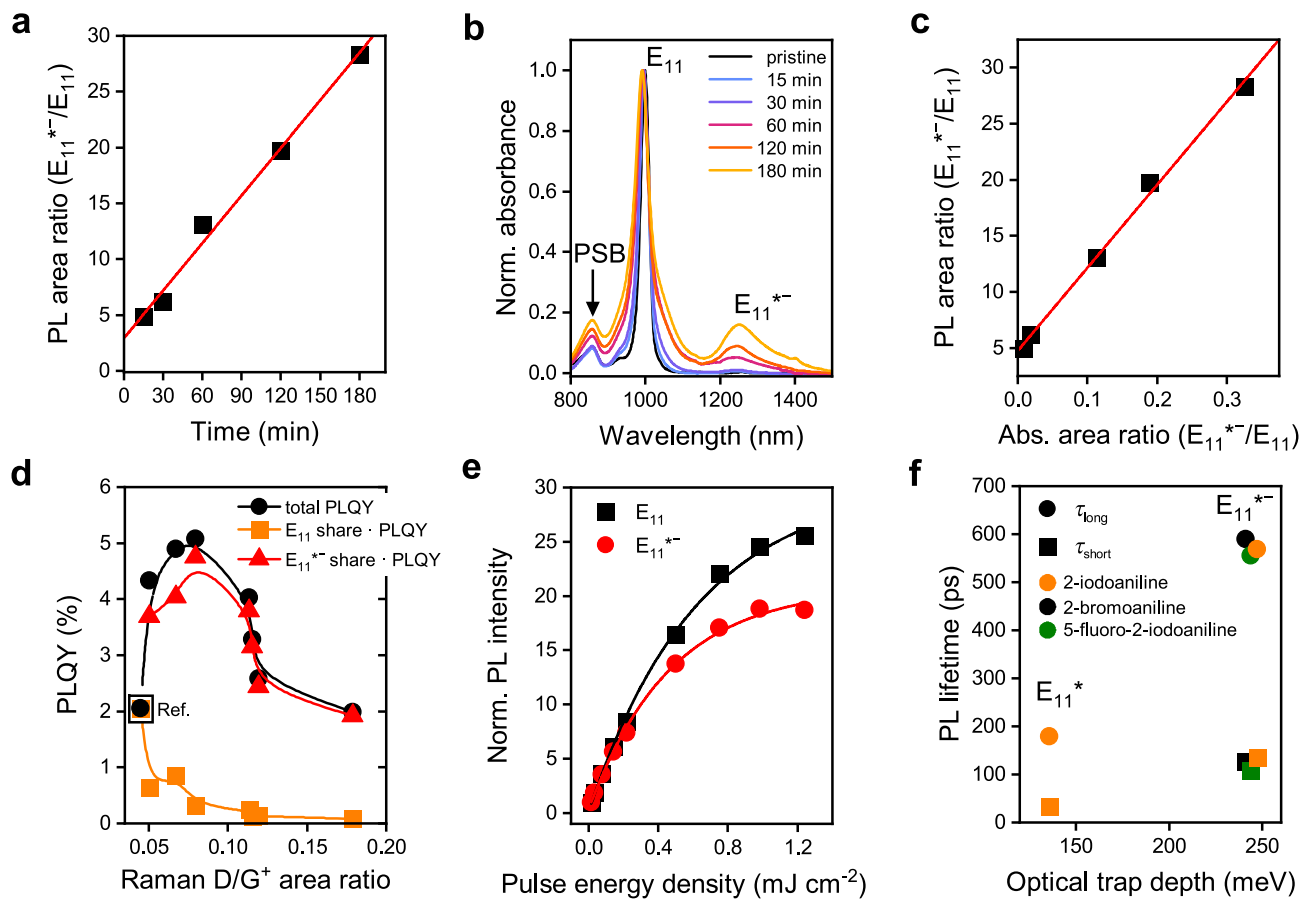

**Fig. 2 Spectroscopic characterization of functionalized (6,5) SWNTs. a–e** (6,5) SWNTs functionalized with 2-iodoaniline in the dark. **a** Integrated $E_{11}^{*-}/E_{11}$ emission ratio vs. reaction time with linear fit. **b** Absorption spectra of functionalized (6,5) SWNTs normalized to $E_{11}$ transition, showing phonon sideband (PSB, ~860 nm) and $E_{11}^{*-}$ absorption band (~1247 nm) increasing with reaction time. It is noteworthy that a broadening of the $E_{11}^{*-}$ defect absorption band may occur due to a scattering background caused by increasing aggregation of SWNTs at very high defect densities. **c** Integrated $E_{11}^{*-}/E_{11}$ emission vs. $E_{11}^{*-}/E_{11}$ absorbance ratio with linear fit. **d** Photoluminescence quantum yield (PLQY) of functionalized SWNTs (total and separated into spectral shares of $E_{11}^{*-}$ and $E_{11}$) vs. integrated Raman D/$G^+$ ratios as metric for defect density (lines are guides to the eye). **e** Normalized intensity of $E_{11}^{*-}$ (red) and $E_{11}$ (black) emission vs. pulse energy density, indicating faster saturation of defect state than band-edge exciton emission. **f** PL lifetime ($\tau_{long}$, $\tau_{short}$) vs. optical trap depth for $E_{11}^{*}$ and $E_{11}^{*-}$ defects created by functionalization of (6,5) SWNTs with 2-iodoaniline (orange), 2-bromoaniline (black), and 5-fluoro-2-iodoaniline (green).

Another characteristic feature of SWNT defect state emission is its nonlinear behavior at high pump power, i.e., the ratio of defect to $E_{11}$ emission depends strongly on the chosen excitation method (lamp or laser, continuous or pulsed) and power (Fig. 2e and Supplementary Fig. 5), and decreases with excitation density. This power dependence can be explained with filling of the long-lived defect states at higher excitation densities[36]. It is more pronounced for the $E_{11}^{*-}$ defects with their emission already saturating at relatively low laser powers compared to the $E_{11}^{*}$ defects (see Supplementary Fig. 6).

The increase in total PLQY and the power dependence of emission are best understood within the framework of the defect state relaxation dynamics. Hence, the PL decays for (6,5) SWNTs functionalized with 2-iodoaniline, 2-bromoaniline, or 5-fluoro-2-iodoaniline were recorded using time-correlated single-photon counting (TCSPC) and fitted as biexponential decays (see Supplementary Fig. 7) with a short ($\tau_{short}$) and long ($\tau_{long}$) lifetime component. As shown in Fig. 2f, the longer lifetimes increase significantly with optical trap depth from 179 ps for $E_{11}^{*}$ (only for 2-iodoaniline, trap depth 136 meV) to up to 589 ps for $E_{11}^{*-}$ (trap depth 241 meV, see also Supplementary Table 2). The values are consistent with previous measurements of $E_{11}^{*-}$ defects that were introduced in much smaller numbers by diazonium chemistry[18,37]. The temperature dependence of the defect emission and thermal detrapping of excitons are presented and discussed in Supplementary Note 2 and Supplementary Figs. 8–11.

**Single-photon emission and individual SWNT spectra**. One important application of luminescent $sp^3$ defects in SWNTs is as quantum-light sources in the near-infrared and at room temperature[18,38]. Defects with large optical trap depths are best suited for single-photon emission at higher temperatures and, ideally, emission should be within one of the standard telecommunication bands (O-band or C-band). The selectively produced $E_{11}^{*-}$ defects in polymer-wrapped SNWTs fullfil a good part of these requirements. To confirm their suitability as single-photon emitters, Hanbury–Brown–Twiss experiments were performed on individualized (6,5) SWNTs with a low concentration of $E_{11}^{*-}$ defects embedded in a polystyrene matrix at room temperature (see Fig. 3a–c). The probability of two consecutive photon detection events is expressed in the second-order photon-correlation function $g^{(2)}(t)$. Under pulsed excitation, the absence of detection events at zero time delay (here, $g^{(2)}(0) = 0.07$) indicates photon antibunching with high single-photon purity (93%) for the $E_{11}^{*-}$ defects. Furthermore, high count rates (~60 kHz) were observed during the measurement with fluctuations near the shot-noise limit.

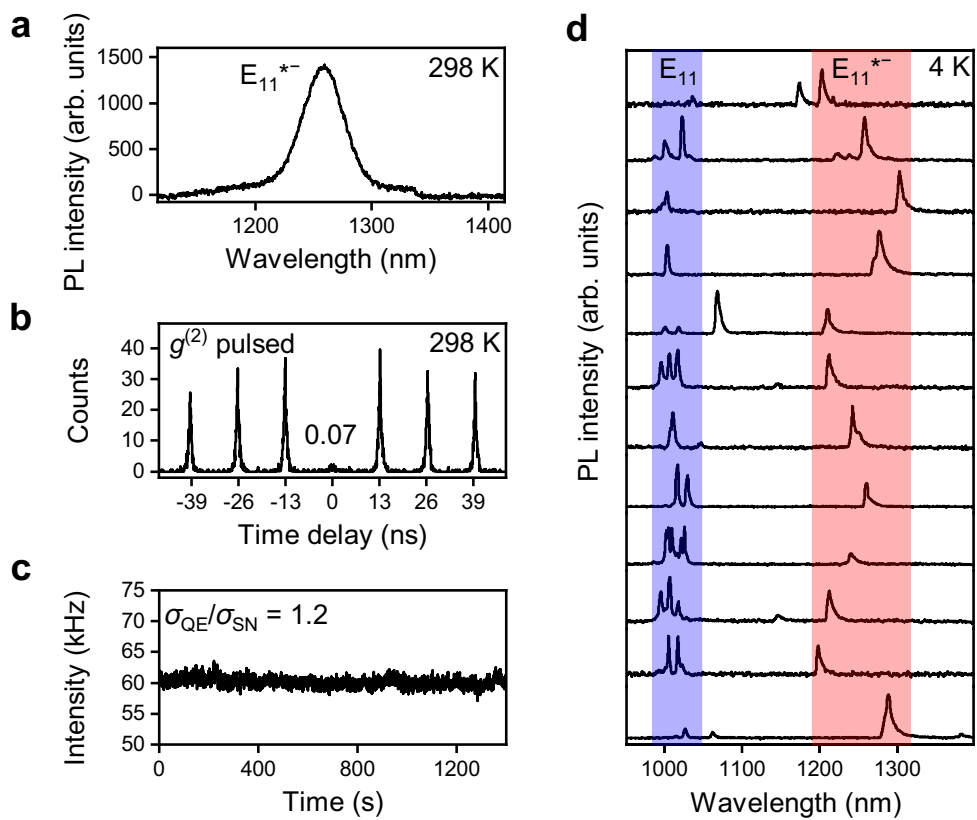

**Fig. 3 PL characteristics of individual (6,5) SWNTs with $E_{11}^{*-}$ defects. a** PL spectrum of a single (6,5) SWNT functionalized with 2-iodoaniline embedded in a polystyrene matrix at room temperature (298 K) collected under continuous wave excitation at the $E_{11}$ transition. **b** Second-order photon-correlation function $g^{(2)}(t)$ of $E_{11}^{*-}$ emission at room temperature, confirming high single-photon purity (93%). **c** PL time trace showing a count rate fluctuation near the shot-noise limit. The deviation of the PL intensity distribution of the emitter ($\sigma_{QE}$) from the Poisson distribution of the detector shot noise ($\sigma_{SN}$) is expressed as $\sigma_{QE}/\sigma_{SN} = ((\langle n^2 \rangle - \langle n \rangle^2)/\langle n \rangle)^{0.5}$, with $n$ being the number of PL counts within a 0.1 s timeframe. **d** PL spectra of 12 individual (6,5) SWNTs functionalized with 2-iodoaniline and embedded in a polystyrene matrix recorded at 4 K. The spectral ranges of the $E_{11}$ (blue) and $E_{11}^{*-}$ (red) emission peaks are highlighted.

Comparable values were found by He et al.[18] at similar emission wavelengths (1280 nm) for polymer-wrapped (6,5) SWNTs functionalized with a diazonium salt. However, in that case, the defect emission of the whole set of individual (6,5) SWNTs was distributed over a large spectral range (1000–1350 nm) due to the presence of many nanotubes with $E_{11}^{*}$ defects and only few with $E_{11}^{*-}$ defects in the ensemble[24]. It is noteworthy that for the application of SWNTs as single-photon sources, homogeneous emission line broadening also represents a significant challenge. However, cavity enhancement of the radiative emission rate via the Purcell effect was recently shown to improve the photon indistinguishability[22]. Here, the exclusive introduction of $E_{11}^{*-}$ defects via the reaction with 2-iodoaniline in the dark enables a better matching with the limited bandwith of such cavities as the distribution of emission wavelengths is reduced to 1200–1300 nm.

The reduced spectral diversity is also reflected at the single-nanotube level as revealed by PL spectra of individualized (6,5) SWNTs with a low $E_{11}^{*-}$ defect density embedded in a polystyrene matrix at 4 K (Fig. 3d). Almost all defect emission peaks appear within an 80 meV window (highlighted in red), with peak widths between 2.9 and 5 meV (limited by a low-resolution grating) and exhibiting the typical asymmetric lineshapes of polymer-wrapped nanotubes. The remaining energetic shifts between emission peaks are likely due to variations in the dielectric environment[18]. It is noteworthy that the segmentation of the $E_{11}$ emission (highlighted in blue) at 4 K originates from random localizations of excitons in long polymer-wrapped SWNTs or from other extrinsic defects introduced during initial processing. They are also present in SWNTs that were not specifically functionalized[39–41]. An integrated spectrum of 62 spots closely resembles the ensemble spectrum of a thin film at room temperature (Supplementary Fig. 12).

**Luminescent $sp^3$ defects in larger diameter SWNTs.** Although the $E_{11}^{*-}$ emission from (6,5) nanotubes (~1250 nm) is already close to relevant telecommunication bands and is well located within the second biological window (1000–1350 nm), further red-shifted PL is desirable. This might be achieved by introducing the same $E_{11}{*}^{-}$ defects in polymer-wrapped semiconducting nanotubes with larger diameters. Two possible candidates are (7,5) SWNTs (diameter 0.83 nm, $E_{11}$ transition at 1047 nm) and (10,5) SWNTs (diameter 1.05 nm, $E_{11}$ transition at 1276 nm). They can be selectively dispersed in toluene using suitable polyfluorene derivatives (poly[(9,9-dioctylfluorenyl-2,7-diyl)] (PFO) and poly[(9,9-dioctylfluorenyl-2,7-diyl)-alt-(1,4-benzo[2,1,3]thiadiazole)] (F8BT), see "Methods" and Supplementary Fig. 13). Nearly monochiral dispersions were functionalized with 2-iodoaniline in the dark. The resulting PL spectra are displayed in Fig. 4a, b, respectively. In both cases, almost only the $E_{11}{*}^{-}$ defect emission was observed; however, the achievable concentration of defects remained relatively low in comparison to the (6,5) SWNTs. Very long reaction times were required to reach moderate $E_{11}{*}^{-}/E_{11}$ PL ratios. The Raman D/G$^+$ ratios for (7,5) nanotubes were also significantly lower than those for (6,5) SWNTs (Supplementary Fig. 14).

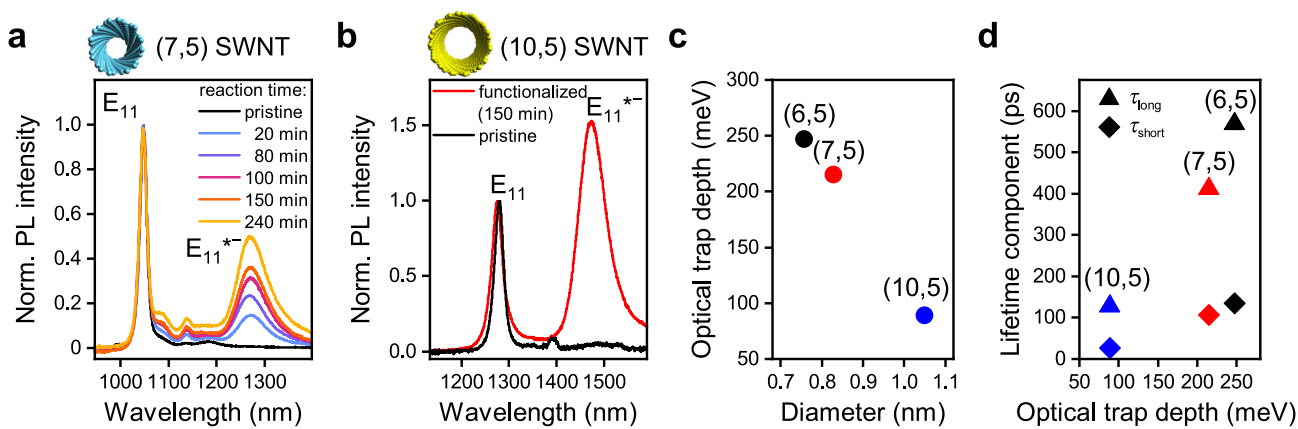

**Fig. 4 Functionalization of polymer-wrapped SWNTs with larger diameters. a** PL spectra of PFO-wrapped (7,5) SWNTs functionalized with 2-iodoaniline in the dark after different reaction times with a red-shifted $E_{11}{*}^{-}$ emission band (~1270 nm). **b** PL spectra of F8BT-wrapped pristine (10,5) SWNTs and functionalized with 2-iodoaniline in the dark for 150 min. Functionalization gives rise to a red-shifted $E_{11}{*}^{-}$ emission band (~1474 nm). The concentration of 2-iodoaniline was kept at 29.30 mmol L$^{-1}$. **c** Optical trap depths of $E_{11}{*}^{-}$ defect states as a function of SWNT diameter. **d** $E_{11}{*}^{-}$ defect fluorescence lifetimes (long and short) depending on optical trap depth.

The $E_{11}$ and $E_{11}{*}^{-}$ emissions from SWNTs with larger diameters occur at longer wavelengths. However, the defect emission bands for (7,5) and (10,5) SWNTs appear at 1270 nm and 1474 nm, respectively, and thus not as far red-shifted as desired for applications. The optical trap depth of the $E_{11}{*}$ defects was previously shown to monotonically decrease with increasing nanotube diameter, making even more red-shifted emission increasingly harder to achieve[42]. This correlation was also found for the $E_{11}{*}^{-}$ defects in (7,5) and (10,5) SWNTs (optical trap depths of 215 meV and 89 meV, respectively) as shown in Fig. 4c. Shallower trap depths also lead to shorter defect PL lifetimes (Fig. 4d) reaching only 128 ps for the long lifetime component of $E_{11}{*}^{-}$ defects in (10,5) nanotubes. Both effects make trapping and radiative recombination at these defect sites less efficient and no increase in PLQY was observed upon functionalization as shown for (7,5) nanotubes with various defect densities (Supplementary Fig. 15).

**Functionalization mechanism and origin of selectivity**. The selectivity of the reaction of 2-haloanilines with SWNTs for defect-binding configurations with either $E_{11}{*}$ or $E_{11}{*}^{-}$ emission depends on conditions such as UV-light illumination and KO$^t$Bu concentration. The different selectivities suggest divergent reaction pathways that are explored in more detail in the following. Previous functionalization reactions of nanotubes with anilines and haloanilines employed UV activation of the reagent or the nanotubes and produced either only $E_{11}{*}$ emission or both $E_{11}{*}$ and $E_{11}{*}^{-}$ emission[23,25,30]. It is commonly assumed that the halogen is eliminated in the process of activation of haloanilines (dehalogenation). However, under the reaction conditions employed here, i.e., with KO$^t$Bu as the base and in the dark, a high reactivity and selectivity was observed not only for reactions with 2-iodoaniline and 2-bromoaniline but also for 2-fluoroaniline (see Fig. 5a). In the latter case, the reaction is not expected to occur via a dehalogenation process, as the C–F bond is one of the strongest known carbon bonds.

To answer the question what chemical group is actually attached to the nanotubes after functionalization, we performed X-ray photoemission spectroscopy (XPS) measurements on thin films of (6,5) SWNTs functionalized with 2-fluoroaniline, using the strong signal of the F 1s core level as a metric (see Fig. 5b). Clearly, the fluorine is present after functionalization, indicating that dehalogenation does not take place. The extracted binding energy (689.5 eV) is in agreement with literature values of alkyl and aryl C–F bonds[43,44]. Additional experiments were carried out with 5-fluoro-2-iodoaniline (Supplementary Fig. 16), with similar results. It is noteworthy that it was not possible to employ thermogravimetric analysis coupled with mass spectrometry[45] to further confirm the nature of attached functional groups due to the necessarily low defect density and limited amount of purified (6,5) SWNTs.

To understand the reaction mechanism of (6,5) SWNTs in the presence of KO$^t$Bu and in the dark, we performed a large number of reference and control experiments (see Supplementary Tables 3 and 4). Surprisingly, functionalization of (6,5) nanotubes and strong defect emission bands ($E_{11}{*}$ and $E_{11}{*}^{-}$) even occurred without any aniline derivative, thus suggesting side reactions that create the same defect-binding configuration. Control reactions were conducted both in toluene and THF. These experiments revealed a complex influence of reaction parameters and only the key aspects will be discussed here (see Supplementary Note 3 for details). Briefly, $E_{11}{*}^{-}$ emission bands only appear in the presence of KO$^t$Bu and a higher selectivity is achieved with the addition of DMSO as a co-solvent. Thus, the introduction of $E_{11}{*}^{-}$ defects can be associated with the basicity of the system, which is greatly increased by DMSO. In contrast to that, $E_{11}{*}$ emission bands appear only in the absence of DMSO or under UV irradiation. Irradiation of polymer-wrapped (6,5) SWNTs in THF or toluene with UV light without reagents did not result in any functionalization. Functionalization with 2-iodoaniline in the dark showed the typical temperature dependence of thermally activated reactions (see Fig. 5c) without losing its selectivity for $E_{11}^{*-}$ defects.

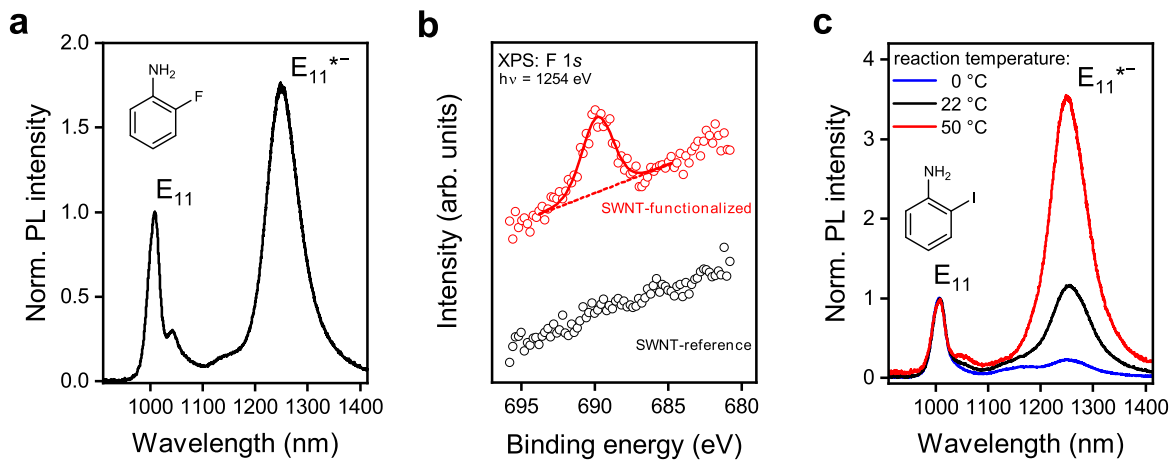

**Fig. 5 Mechanistic investigation of functionalization. a** PL spectrum of (6,5) SWNTs functionalized with 2-fluoroaniline in the dark with a strong defect emission band at 1250 nm. **b** F 1s XPS spectra of (6,5) SWNTs after reaction with 2-fluoroaniline with (red circles including peak fit, red line) and without (black circles; reference) addition of KO$^t$Bu as base. The presence of the F 1s signal indicates covalent functionalization with 2-fluoroaniline, whereas the absence of this signal for the reference sample indicates lack of functionalization. **c** PL spectra of (6,5) SWNTs functionalized with 2-iodoaniline at different reaction temperatures in the dark. The concentrations of 2-fluoroaniline and 2-iodoaniline were kept at 29.30 mmol L$^{-1}$.

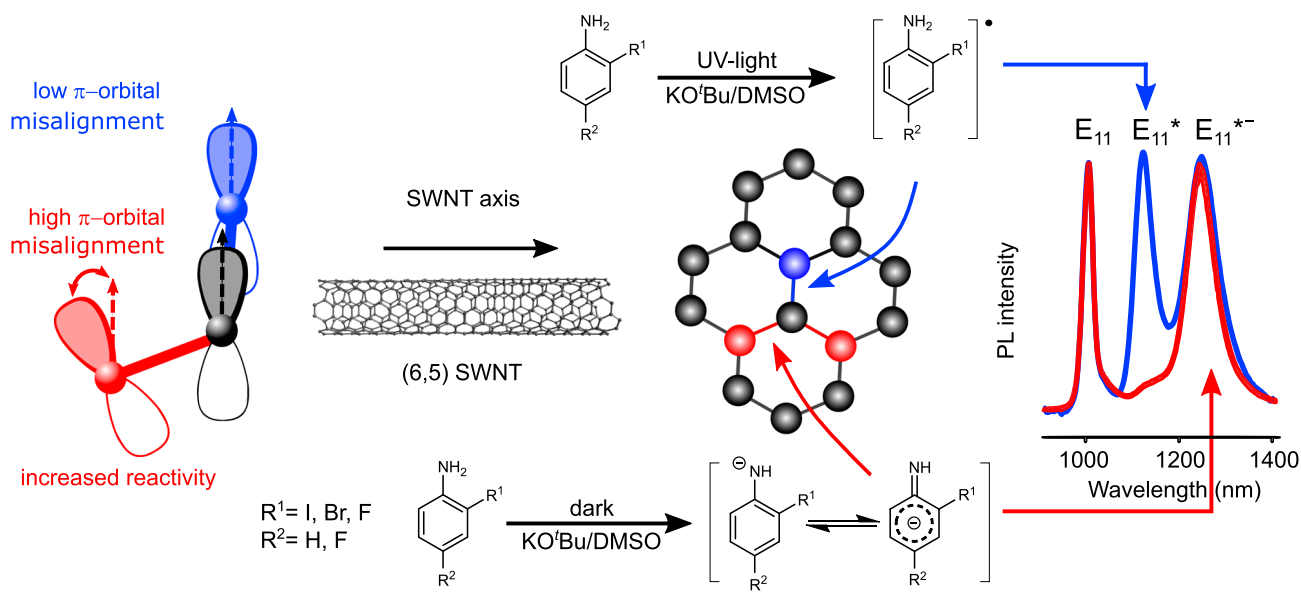

**Fig. 6 Proposed functionalization mechanism for (6,5) SWNTs.** Aniline derivatives can follow two reaction paths in the presence of KO$^t$Bu and DMSO. Under UV-light irradiation, a radical species is formed that attacks the C–C bonds in circumferential direction, leading to the defect configuration for the $E_{11}^*$ emission band (blue line). When the reaction is performed in the dark, deprotonation of the amine group and charge delocalization occurs. A nucleophilic attack on bonds with large $\pi$-orbital misalignment creates the defect configuration for the $E_{11}^{*-}$ emission band (red line). Similar reaction paths are expected for phenols, thiophenols, and indoles with different carbanion intermediates. It is noteworthy that the deprotonated aniline species is in equilibrium with its protonated form.

Considering the available data, we propose two possible functionalization mechanisms based on radical and nucleophilic reaction paths that determine the final defect-binding configurations. The $E_{11}^*$ emission band is associated with the ortho-L$_{90}$ (ortho++) configuration[20,24,27], which is oriented in circumferential direction for (6,5) SWNTs (see Fig. 6, blue carbon atom). It is usually observed for reactions with diazonium salts and other monovalent functionalization reactions, e.g., with aryl halides, relying on radical formation. In our case, $E_{11}^*$ emission mainly appears when the formation of radicals is assisted by UV-light illumination or electron transfer. Consequently, we assign the $E_{11}^*$ emission to a reaction path where the carbon bonds in circumferential direction of the SWNT are attacked by radical species (indicated in blue in Fig. 6). Recent studies have shown that functionalization by radicals can also introduce further red-shifted $E_{11}^{*-}$ defects, but only when the already existing defects induce a higher $\pi$-orbital misalignment in bonds along the tube axis[29]. In contrast to that, non-radical reactive species exhibit an intrinsic preference for C–C bonds with large $\pi$-orbital misalignment and thus preferably attack carbon bonds in the axial direction (highlighted in red). Such reactive species are nucleophilic intermediates.

Nucleophilic intermediates can be formed via deprotonation of aniline derivatives by a base and stabilization of the anion through various possible resonance structures[46,47]. This initial deprotonation represents an equilibrium reaction, which shifts depending on the concentration of the base in relation to the aniline reagent. The observed increase in $E_{11}^{*-}$ emission for higher concentrations of base can therefore be attributed to a shift in favor of the deprotonated aniline species (see Fig. 1b). Nucleophilic attack of the SWNT would then lead to a charged SWNT intermediate that must be saturated quickly, e.g., by a proton from the solvent, yielding the defect configuration for $E_{11}^{*-}$ emission (i.e., ortho-$L_{30}$). Similar nucleophilic reaction mechanisms have been reported for the functionalization of $C_{60}$-fullerenes via KO$^t$Bu-promoted coupling of indoles[48] and phenols[49]. In agreement with the proposed mechanism, strong $E_{11}^{*-}$ emission (Supplementary Fig. 17 and Supplementary Table 5) was also observed for the functionalization of (6,5) SWNTs with indole, 2-iodophenol, and thiophenol, although some other emission features were introduced as well. Significant variations of the delocalization of charge can be expected for different anilines, as well as phenols and indoles, thus leading to differences in stabilization and, consequently, reactivity of the carbanion intermediate. Depending on the reactivity of the carbanion intermediate, different selectivities of $E_{11}^{*-}$ vs. $E_{11}^*$ defects may be achieved (see Supplementary Note 4). Under inert conditions, a strong increase in reactivity is observed (Supplementary Fig. 18), indicating that oxidation of the expected negatively charged intermediates (e.g., SWNT or other carbanionic intermediates, see Supplementary Note 5) and quenching of the base by moisture are inhibiting the functionalization under ambient conditions to some degree. It is also noteworthy that some nucleophilic addition reactions with nanotubes have been known for over a decade[2]; however, they usually involve very strong organic bases such as sodium hydride[50] or organolithium/organomagnesium[51,52] compounds. They have never been applied to introduce luminescent $sp^3$ defects at low concentrations and thus their selectivity for certain binding configurations is not known.

## Discussion

We have demonstrated a facile method for the functionalization of semiconducting SWNTs that selectively and exclusively creates luminescent $sp^3$ defects with a binding configuration for strongly red-shifted $E_{11}^{*-}$ emission instead of the commonly found $E_{11}^*$ emission. The reaction can be performed with various 2-haloanilines and other aniline derivatives at room temperature and in air. The final defect density is precisely and easily tuned by the concentration of the base KO$^t$Bu and the reaction time. Optimized defect densities enhance the absolute PLQYs to up to 5% for polymer-wrapped (6,5) SWNTs. The deep optical trap depth and long fluorescence lifetimes of the $E_{11}^{*-}$ defects in (6,5) SWNTs enable their application as near-infrared single-photon emitters at room temperature with high single-photon purity. Although the defect emission wavelength of 1250 nm lies well within the second biological window and thus might be useful for in vivo imaging, emission even further in the near-infrared was achieved by functionalization of (7,5) and (10,5) nanotubes. The high selectivity of this base-mediated reaction for a specific defect-binding configuration leading to the $E_{11}^{*-}$ emission probably relies on a nucleophilic reaction mechanism in contrast to commonly found aryl radical and reductive alkylation reactions. This type of base-mediated nucleophilic functionalization expands and complements the available chemistry for the controlled introduction of luminescent $sp^3$ defects in SWNTs and opens the path to a broader variety of functional groups and thus applications.

## Methods

**Polymer-sorting of SWNTs.** The (6,5) SWNTs were selectively dispersed from CoMoCAT raw material (Chasm SG65i-L63) by shear-force mixing (Silverson L2/Air, 10,230 r.p.m., 20 °C, 72 h) in a solution of PFO-BPy (American Dye Source, $M_w$ = 40 kg mol$^{-1}$, 0.5 g L$^{-1}$) in toluene. The (7,5) SWNTs were sorted by dispersion in toluene with PFO (American Dye Source, $M_w$ > 20 kg mol$^{-1}$, 0.9 g L$^{-1}$) as the wrapping polymer. For the enrichment of (10,5) SWNTs, HiPco raw material (Unidym, Inc., Batch No. 2172) was dispersed by tip sonication with F8BT (American Dye Source, $M_W$ = 59 kg mol$^{-1}$, 4 g L$^{-1}$) in toluene. Unexfoliated material was removed by centrifugation at 60,000 × g (Beckman Coulter Avanti J26XP centrifuge) and filtration (polytetrafluoroethylene (PTFE) syringe filter, 5 μm pore size). To remove excess polymer, the SWNTs were collected by vacuum filtration through a PTFE membrane filter (0.1 μm pore size). Filter cakes were washed three times with toluene at 80 °C and redispersed in fresh toluene by bath sonication. Excess F8BT was removed from (10,5) SWNT dispersions by ultra-centrifugation (Optima XPN-80 centrifuge, 284,600 × g, 24 h). The polymer-rich supernatant was discarded and the obtained SWNT pellets were washed with THF three times before redispersion in fresh toluene.

**Introduction of $sp^3$ defects.** Polymer-wrapped SWNTs were functionalized with a range of commercially available aryl compounds (2-iodoaniline, 2-bromoaniline, 2-fluoroaniline, 5-fluoro-2-iodoaniline, 2-iodophenol, thiophenol, indole; used as received from Sigma Aldrich, ≥97%). For a step-by-step protocol, refer to Supplementary Methods 1. Briefly, an appropriate amount of aryl reagent was dissolved in toluene to achieve a final reaction concentration of 29.30 mmol L$^{-1}$. DMSO (anhydrous) and KO$^t$Bu (Sigma Aldrich, 98%) dissolved in THF (anhydrous) were added to this solution. Finally, an enriched SWNT dispersion in toluene was added to the mixture such that the SWNT concentration in the reaction mixture corresponded to an optical density of 0.3 cm$^{-1}$ at the $E_{11}$ transition. The final solvent composition was 83.3 : 8.3 : 8.3 vol-% toluene/DMSO/THF. Functionalization was performed in the dark or under illumination with UV-light (LED SOLIS-365C, Thorlabs, 365 nm, 1.9 mW mm$^{-2}$) at room temperature. The reaction was stopped after a given time by vacuum filtration of the reaction mixture through a PTFE membrane filter (0.1 μm pore size). The collected SWNTs were washed with methanol and toluene to remove unreacted reagents and side products. The resulting filter cake was redispersed by bath sonication in toluene with fresh wrapping polymer for stabilization.

**Spectroscopic characterization.** Absorption spectra were acquired with a Cary 6000i UV-VIS-NIR spectrophotometer (Varian, Inc.). PL spectra of dispersions at low excitation power and controlled temperature were recorded using a Fluorolog spectrofluorometer (HORIBA) with a 450 W xenon arc lamp and liquid-nitrogen-cooled InGaAs line camera. For acquisition of room temperature, PL spectra of dispersions, and time-correlated single-photon counting, the wavelength-tunable output of a picosecond-pulsed (~6 ps pulse width) supercontinuum laser (Fianium WhiteLase SC400) was used to excite SWNTs at the $E_{22}$ transition. PL spectra were collected with an Acton SpectraPro SP2358 spectrograph (grating blaze 1200 nm, 150 lines mm$^{-1}$) and a liquid-nitrogen-cooled InGaAs line camera (Princeton Instruments, OMA-V:1024). Wavelength-dependent PL lifetimes were measured by focusing the spectrally filtered emission onto a gated InGaAs/InP avalanche photodiode (Micro Photon Devices). The absolute PLQY was determined using an integration sphere (see Supplementary Methods 1 for details).

For low-temperature single-nanotube spectroscopy and temperature-dependent measurements, a dispersion of functionalized SWNTs was diluted with a toluene solution of polystyrene (Polymer Source, Inc., $M_w$ = 230 kg mol$^{-1}$) and spin-coated onto glass slides coated with 150 nm of gold. Samples were mounted in a closed-cycle liquid helium optical cryostat (Montana Instruments Cryostation s50) and excited with a continuous wave laser diode (OBIS, Coherent, Inc., 640 nm). PL images and spectra were recorded with a thermoelectrically cooled InGaAs camera (NIRvana 640ST) coupled to an IsoPlane SCT-320 spectrograph (Princeton Instruments). A standard Hanbury–Brown and Twiss setup was employed for measurements of second-order photon-correlation function $g^{(2)}(t)$ of individual SWNTs at room temperature. The sample was excited with a supercontinuum laser (NKT Photonics, SuperK EXTREME, 6 ps pulse width, 78 MHz repetition rate, tuned to 995 nm) and the emission detected with a pair of superconducting single-photon detectors (Scontel, TCOPRS-CCR-SW-85). Detection events were recorded and correlated using a TCSPC module (PicoQuant, PicoHarp300).

**X-ray photoemission spectroscopy.** Samples were prepared by drop casting of highly functionalized (6,5) SWNT dispersions on gold-coated silicon substrates. XPS measurements were performed with a MAX 200 (Leybold–Heraeus) spectrometer equipped with an Mg Kα X-ray source (260 W; ~1.5 cm distance to the samples) and a hemispherical analyzer (EA 200; Leybold–Heraeus). Spectra were acquired in normal emission geometry with an energy resolution of ~0.9 eV. The binding energy scale was calibrated to the Au $4f_{7/2}$ line of the substrate at 84.0 eV. Potential damage induced by X-rays was kept as low as possible. The measurements were carried out under ultra-high vacuum conditions at a base pressure of ~3 × 10$^{-9}$ mbar.

## Data availability

The datasets generated and/or analyzed during the current study are available in the heiDATA repository: https://doi.org/10.11588/data/RI2KLV.

## Acknowledgements
This project has received funding from the European Research Council (ERC) under the European Union's Horizon 2020 research and innovation program (Grant agreement number 817494 "TRIFECTs"). S.Z. acknowledges funding from the Alexander von Humboldt Foundation and A.H. from the European Research Council (ERC) under the Grant agreement number 772195 and the Deutsche Forschungsgemeinschaft (DFG, German Research Foundation) under Germany's Excellence Strategy EXC-2111-390814868.

### Author contributions
S.S., F.J.B., and S.L. performed synthesis and characterization. N.F.Z. and A.Y. contributed low-temperature spectroscopy. S.Z. and A.H. performed photon-correlation spectroscopy measurements. A.A. and M.Z. provided photoemission spectroscopy measurements. J.Z. conceived and supervised the project. S.S. and J.Z. wrote the manuscript. All authors discussed the data analysis and commented on the manuscript.

### Funding
Open Access funding enabled and organized by Projekt DEAL.

### Competing interests
The authors declare no competing interests.

### Additional information
**Supplementary information** The online version contains supplementary material available at https://doi.org/10.1038/s41467-021-22307-9.

**Correspondence** and requests for materials should be addressed to J.Z.

**Peer review information** *Nature Communications* thanks Dawid Janas and the other, anonymous, reviewer(s) for their contribution to the peer review of this work. Peer reviewer reports are available.

**Reprints and permission information** is available at http://www.nature.com/reprints

**Publisher's note** Springer Nature remains neutral with regard to jurisdictional claims in published maps and institutional affiliations.

# Synthetic control over the binding configuration of luminescent $sp^3$-defects in single-walled carbon nanotubes

*Simon Settele[1], Felix J. Berger[1,2], Sebastian Lindenthal[1], Shen Zhao[3], Abdurrahman Ali El Yumin[1,2], Nicolas F. Zorn,[1,2], Andika Asyuda[1], Michael Zharnikov[1], Alexander Högele[3,4] and Jana Zaumseil[1,2]\**

[1] Institute for Physical Chemistry, Universität Heidelberg, D-69120 Heidelberg, Germany

[2] Centre for Advanced Materials, Universität Heidelberg, D-69120 Heidelberg, Germany

[3] Fakultät für Physik, Munich Quantum Center, and Center for NanoScience (CeNS), Ludwig-Maximilians-Universität München, D-80539 München, Germany

[4] Munich Center for Quantum Science and Technology (MCQST), D-80799 München, Germany

\* corresponding author: *zaumseil@uni-heidelberg.de*

# CONTENTS

# Supplementary Methods 1

**Raman Spectra of Functionalized (6,5) SWNTs**

(Functionalized) single-walled carbon nanotubes (SWNTs) dispersions were drop-cast on glass substrates and Raman spectra were recorded using a Renishaw inVia Reflex confocal Raman microscope with a 532 nm laser for near-resonant excitation. More than 1000 spectra per sample were collected, averaged and baseline-corrected.

**Fluorescence Lifetime Measurements (Time-Correlated Single Photon Counting, TCSPC)**

Fluorescence lifetimes of luminescent defect states were determined by time-correlated single photon counting (TCSPC) as reported previously.[1] The wavelength-tunable output of a picosecond-pulsed (~6 ps pulse width) supercontinuum laser source (Fianium WhiteLase SC400) focused into a nanotube dispersion via a 50x NIR-optimized objective (N.A. 0.65, Olympus) was used to excite SWNTs at the $E_{22}$ transition (e.g., 575 nm for (6,5) SWNTs). Emitted photons were collected with the same objective. The photoluminescence was filtered by a spectrograph (Acton SpectraPro SP2358, grating blaze 1200 nm, 150 lines mm$^{-1}$) and focused onto a gated InGaAs/InP avalanche photodiode (Micro Photon Devices). Arrival times of the detected photons were recorded with a time-correlated single-photon counting module (Picoharp 300, Picoquant GmbH). The instrument response function (IRF) was determined by the fast, instrument-limited photoluminescence decay at the $E_{11}$ transition (e.g. ~1000 nm for (6,5) SWNTs). All fluorescence decay histograms were fitted with a biexponential model in a reconvolution procedure.

**Normalization of Photoluminescence Spectra**

To display changes in defect density (see Figure 1b-d and Figure 4a,b of the main manuscript) normalized PL spectra are presented instead of absolute emission intensities for the following experimental reasons:

(1) To stop the reaction and remove side products, the functionalization protocol includes a filtration, washing and redispersion sequence (see detailed *sp*$^3$ functionalization protocol). It is not possible to simply record the PL spectra during the reaction. As with pristine SWNTs, the yield of the redispersion step varies and depends on parameters such as sonication power, environmental humidity and temperature. Hence, the resulting dispersions have different concentrations of the dispersed SWNTs and consequently different absolute PL intensities.

(2) While the absorption spectrum of the SWNTs is unaffected at lower levels of functionalization, samples with higher defect densities display a significant reduction of the main absorption band. Hence, the effective absorption cross-section for the functionalized SWNTs also depends on defect density preventing an accurate correction for the yield of the redispersion step.

(3) PL measurements were performed by focusing the excitation laser into a cuvette through an objective. This configuration has the advantage that the near-infrared absorption of toluene does not affect the PL spectra due to the extremely short path length within the liquid. However, even slight differences in the quality of the focus have an impact on the absolute PL intensities, while the spectrum is usually unaffected by minor changes in focus.

As a result, comparison of the absolute PL intensities across a sample series is unreliable and only normalized PL spectra are shown. However, direct and reliable values of the emission efficiencies are provided by photoluminescence quantum yield measurements (see below and Figure 2d of the main manuscript).

**Photoluminescence Quantum Yield Measurements**

The absolute photoluminescence quantum yield (PLQY) of pristine and functionalized nanotubes in dispersion was determined using an integrating sphere.[1,2] The SWNT dispersions were adjusted to an optical density of < 0.2 cm$^{-1}$ at the $E_{11}$ transition and placed in the center of an integrating sphere (LabSphere, Spectralon coating). A laser beam tuned to the $E_{22}$ transition was directed onto the sample and the signal (scattered laser light and photoluminescence) was transmitted to the spectrometer via an optical fiber. To account for absorption of the solvent at the excitation wavelength, the same measurement was repeated with the pure solvent (toluene). PQLY was calculated as the ratio of emitted to absorbed photons. The wavelength-dependent detection efficiency and losses were corrected by recording a reference spectrum of a stabilized tungsten halogen light source with known spectral power distribution (Thorlabs SLS201/M, 300-2600 nm) that was placed in front of the integration sphere.

**Low and Variable Temperature Spectroscopy of SWNTs**

Low-temperature photoluminescence spectra of individual, PFO-BPy-wrapped, functionalized (6,5) SWNTs embedded in polystyrene and dense films of nanotubes were recorded using a closed-cycle liquid helium optical cryostat (Montana Instruments Cryostation s50) with an adjustable temperature between 3.8 K and 300 K. The nanotube samples were excited with a continuous wave laser diode (OBIS, Coherent Inc., 640 nm) through an infrared 50x long working distance objective (Mitutoyo, N.A. = 0.42) mounted outside the cryostat. The laser power was typically ~100 µW and the polarization was adjusted with a $\lambda/2$ plate to match the orientation of individual nanotubes. PL spectra were acquired with a thermoelectrically cooled InGaAs camera (NIRvana 640ST, Princeton Instruments) mounted on a corresponding spectrograph (IsoPlane SCT-320, Princeton Instruments).

**Room Temperature Spectroscopy of Individual SWNTs and Autocorrelation Measurements**

Room-temperature photoluminescence and autocorrelation measurements were performed on individual functionalized (6,5) SWNTs embedded in polystyrene with a home-built confocal microscope with slip-stick positioners (ANPxy101 and ANPz102, attocube systems). A wavelength-tunable Ti:sapphire laser (Mira, Coherent) in continuous wave mode served as the excitation source and was tuned to 995 nm to be in resonance with the $E_{11}$ transition of (6,5) SWNTs. The excitation laser was focused onto the sample with an apochromatic objective (LT-APO/IR/0.81, attocube systems) and PL was collected by the same objective and spectrally filtered using a tunable long-pass filter (TLP01-1116, Semrock). A spectrometer (Acton SP2500, Roper Scientific) coupled with a liquid-nitrogen cooled InGaAs camera (OMA V: 1024-1.7, Roper Scientific) were used to record photoluminescence spectra. For time-resolved photoluminescence and pulsed photon correlation in a standard Hanbury-Brown and Twiss setup, the sample was excited using a supercontinuum laser (SuperK EXTREME, NKT Photonics) with a 6 ps pulse width and a 78 MHz repetition rate, tuned to 995 nm by a set of spectral filters. Emission was directed onto a superconducting single photon detector (TCOPRS-CCR-SW-85, Scontel) and photon detection events were recorded using a time-correlated single-photon counting module (PicoHarp300, PicoQuant).

# Detailed *sp*$^3$ Functionalization Protocol

**General Remarks**

Dispersions of polymer-wrapped SWNTs were functionalized with *sp*$^3$-defects via a potassium *tert*-butoxide (KO$^t$Bu)-mediated coupling approach using primarily 2-haloanilines (later also indoles/thiols/phenols) in a toluene/dimethylsulfoxid (DMSO)/tetrahydrofuran (THF) mixture. While the concentrations of the reactive reagent (e.g., aniline derivative) and DMSO were kept constant, the degree of functionalization could be controlled by the amount of base (KO$^t$Bu), reaction time or temperature. SWNT dispersions (after removal of excess polymer) were used as the starting material and the nanotube concentration was always kept at 0.54 mg L$^{-1}$ (corresponding to an E$_{11}$ absorbance of 0.3 cm$^{-1}$ for (6,5) SWNTs).[3]

Filtration of the nanotube dispersion and redispersion in pure solvent is used to remove excess wrapping polymer and was performed for all reactions presented in this study for higher reproducibility. However, this step is mainly needed for reactions with UV irradiation to avoid light attenuation at 365 nm due to strong absorption by the polymer. The filtration step is not necessary when the functionalization is performed in the dark and the concentration of the wrapping polymer is below ~0.3 g L$^{-1}$.

**Reagents**

All chemicals used in the described functionalization reactions were purchased from Sigma Aldrich and used without further purification:

2-iodoaniline (98%), 2-bromoaniline (97%), 2-fluoroaniline (99%), 5-fluoro-2-iodoaniline (97%), 2-iodophenol (98%), thiophenol (97%), indole (99%), potassium *tert*-butoxide (98%), dimethylsulfoxide (anhydrous, ≥99.9%), tetrahydrofuran (anhydrous ≥99.9% inhibitor free).

The quality of DMSO was found to have a significant impact on the reactivity but not on the selectivity. In order to achieve comparable reactivities as presented in this work, freshly dried DMSO is recommended. Stock solutions of KO$^t$Bu and DMSO should be stored under inert gas atmosphere, but the functionalization itself can be performed in an open flask at room temperature.

**Step by Step Reaction Protocol**

1. Filter polymer-wrapped SWNTs dispersion over a PTFE membrane (Merck Millipore JVWP, 0.1 μm pore size) and wash the resulting filter cake three times (each 5 minutes) with hot toluene (80 °C) to remove excess wrapping polymer.
2. Redisperse the washed filter cake by ultrasonication and adjust the SWNT concentration to an optical density of >1.0 cm$^{-1}$ at the E$_{11}$ absorption peak.
3. Prepare a solution of KO$^t$Bu in dry THF under nitrogen atmosphere. The total volume of THF should result in 8.3 vol-% of the final reaction volume. The amount of KO$^t$Bu used in this work is always given as molar equivalents (eq.) with respect to the used aryl reagent (e.g., aniline derivative). See Supplementary Table 1 for typical values.
4. Dissolve desired amount of reagent (e.g., aniline derivative) in a clear glass vial equipped with a stirring bar with appropriate amount of toluene. Note: After the addition of all components the final concentration of the reagent should be 29.30 mmol L$^{-1}$.
5. Add dry DMSO to the solution of the aryl compound in toluene to obtain 8.3 vol-% of DMSO in the final mixture.
6. Add the prepared KO$^t$Bu/THF solution.
7. Add the prepared (polymer-depleted) SWNT dispersion such that the final reaction mixture has an optical density at the E$_{11}$ transition of 0.3 cm$^{-1}$.
8. Mix the reaction volume thoroughly. The final ratio of toluene/DMSO/THF should be 83.3 : 8.3 : 8.3 vol-%.

It is important that the reagent (e.g., aniline derivative) and DMSO are present in the reaction mixture before the addition of KO$^t$Bu/THF to prevent undesired side-reactions.

*When functionalization is performed in the dark:*

9. Protect the glass vial from light and stir the reaction mixture for the desired duration at room temperature. Reaction times can range between 15 min and 180 min.

*When functionalization is performed under UV light irradiation:*

10. Irradiate the glass vial with UV-light (365 nm, here SOLIS-365C, Thorlabs, 1.9 mW/mm$^2$) under continuous stirring. Reaction times usually vary between 10 min and 45 min.

*Work up:*

11. After the desired functionalization time has elapsed, pass the reaction mixture through a PTFE membrane (e.g. Merck Millipore JVWP, 0.1 mm pore size) and wash the filter cake with approximately 5 mL MeOH and 5 mL toluene on the filtration setup.
12. Redisperse the filter cake in the desired amount of toluene with a low concentration of fresh wrapping polymer (e.g. 0.1 g L$^{-1}$) by bath sonication for 20 min. Addition of wrapping polymer is not strictly necessary, but increases the colloidal stability of the dispersion for characterization. Note that for high defect densities (D/G$^+$ ratio greater than 0.2) the yield of the redispersion process starts to decline due to increasing aggregation of the functionalized SWNTs.

**Example Values for Reaction Mixture**

**Supplementary Table 1:** Exemplary values for functionalization of PFO-BPy wrapped (6,5) SWNTs with 2-iodoaniline and 2 eq. KO$^t$Bu as performed in this study.

| Reagent | Amount |
|---|---|
| KO$^t$Bu (2 eq.) | 78.5 mg |
| THF | 1 mL |
| 2-iodoaniline | 76.6 mg |
| Toluene | 7.188 mL |
| DMSO | 1 mL |
| Polymer-free (6,5) SWNTs (1.28 cm$^{-1}$ at the E$_{11}$) in toluene | 2.812 mL |
| **Total reaction volume** | **12 mL** |

**Spectroscopic Characterization of Pristine PFO-BPy Wrapped (6,5) SWNTs**

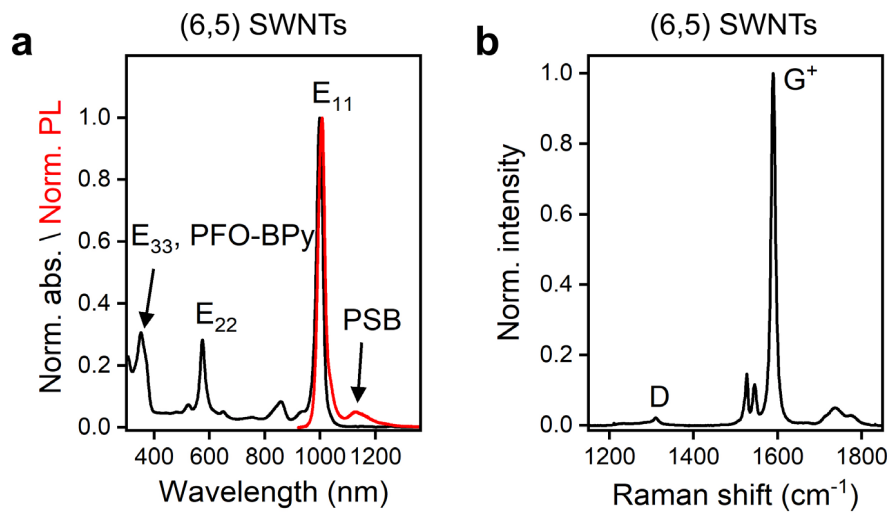

**Supplementary Figure 1. a**, Normalized absorption (black) and photoluminescence (red) spectra showing characteristic excitonic transitions of (6,5) SWNTs labelled as $E_{11}$, $E_{22}$ and $E_{33}$ and PSB (photoluminescence sideband). The small absorption peak below 400 nm indicates the low residual concentration of PFO-BPy after removal of excess wrapping polymer. **b**, Normalized and averaged Raman spectra (resonant excitation at 532 nm) of drop-cast pristine (6,5) SWNTs showing the G-modes ($G^+$, $G^-$) and D-mode. The G modes correspond to the longitudinal and tangential phonons of the hexagonal $sp^2$-carbon lattice, while the D-mode is related to disorder and defects. The ratio of their intensities (D/$G^+$) is used as a metric for the defect density.

## Supplementary Note 1: Concentration Effects

Following the general procedure described above, (6,5) SWNTs were also functionalized using different concentrations of 2-iodoaniline and KO$^t$Bu. The ratio of 2-iodoaniline to base was kept at 1:2. As the the functionalization was mostly conducted with a large excess of reagent (58.6 – 14.65 mmoL L$^{-1}$ 2-iodoaniline) compared to the SWNTs small changes in concentration did not significantly alter the functionalization process as shown in Supplementary Figure 2. For much lower concentrations (<10 mmoL L$^{-1}$ 2-iodoaniline), the E$_{11}$*$^-$ emission feature drops significantly. This is expected, because (1) the amount of reactive intermediate is greatly reduced and (2) the relative quenching of KO$^t$Bu by moisture may be increased as we perform the reaction in an open flask. At higher concentrations this quenching becomes negligible.

To further understand the effect of the concentration of 2-iodoaniline, we conducted functionalization with various concentrations of 2-iodoaniline, while the concentration of KO$^t$Bu was kept constant. This led to an increase of the ratio of 2-iodoaniline to KO$^t$Bu. Since the functionalization is still performed with large excess of the reactant compared to the SWNTs, the E$_{11}$*$^-$ emission is higher for higher 2-iodoaniline/KO$^t$Bu ratios in agreement with Figure 1c.

**Concentration Effects on the Functionalization Process**

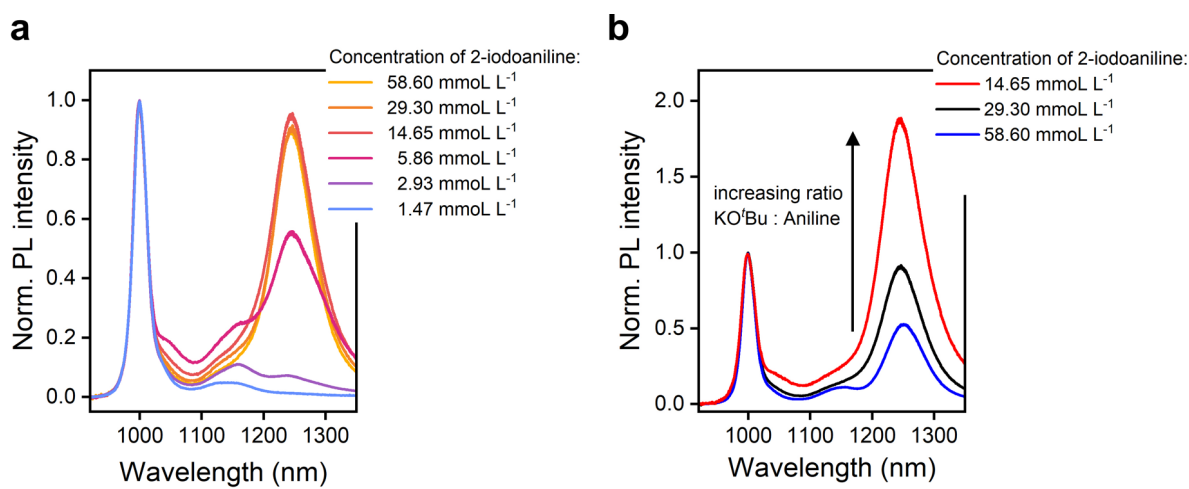

**Supplementary Figure 2. a,** Normalized photoluminescence spectra of (6,5) SWNTs after functionalization with different concentrations of 2-iodoaniline and 2 eq. of KO$^t$Bu for 10 minutes in the dark in toluene/DMSO/THF. **b,** Normalized photoluminescence spectra of (6,5) SWNTs after functionalization with different concentrations of 2-iodoaniline for 10 minutes in the dark in toluene/DMSO/THF. The concentration of KO$^t$Bu was kept at 58.60 mmol L$^{-1}$.

# Functionalization with Different Aniline Derivatives

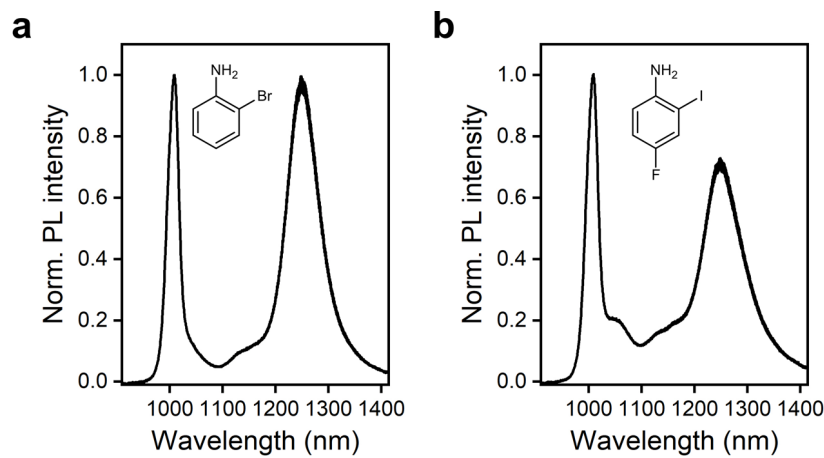

**Supplementary Figure 3. a,b**, Normalized photoluminescence spectra of (6,5) SWNTs after functionalization with 2-bromoaniline (**a**) and 5-fluoro-2-iodoaniline (**b**) with 2 eq. of KO$^t$Bu for 30 minutes in the dark in toluene/DMSO/THF. The concentration of 2-bromoaniline and 5-fluoro-2-iodoaniline was kept constant at 29.30 mmol L$^{-1}$.

# Raman Spectra of Functionalized (6,5) SWNTs

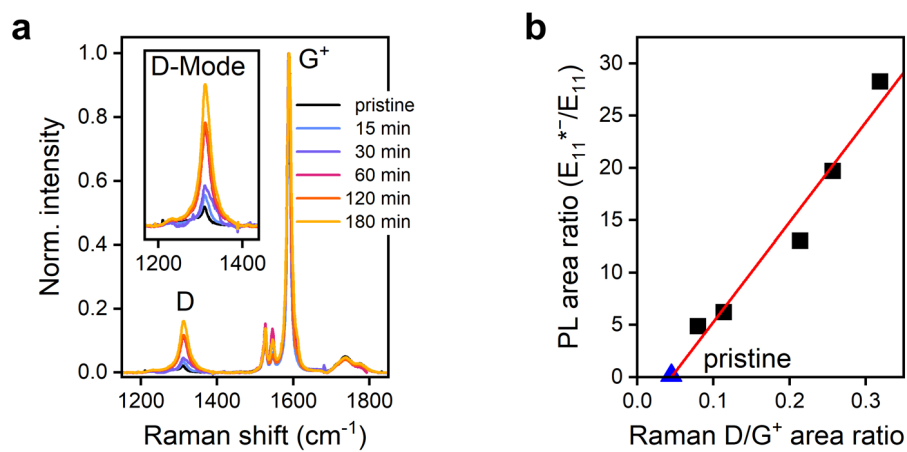

**Supplementary Figure 4. a**, Averaged Raman spectra of (6,5) SWNTs functionalized with 2-iodoaniline and KO$^t$Bu in the dark (resulting in $E_{11}$*$^-$ defects) at different reaction times (corresponding to data in Figure 1d) normalized to G$^+$-mode. The inset shows the D-mode region. The defect-related D-mode intensity increases with reaction time. **b**, Integrated $E_{11}$*$^-$/$E_{11}$ emission ratios vs. integrated Raman D/G$^+$ ratios and linear fit as metric of defect density. Blue triangle: data point for pristine (6,5) SWNTs (see Supplementary Figure 1b).

**Power-Dependence of $E_{11}^{*-}$ Defect State Photoluminescence**

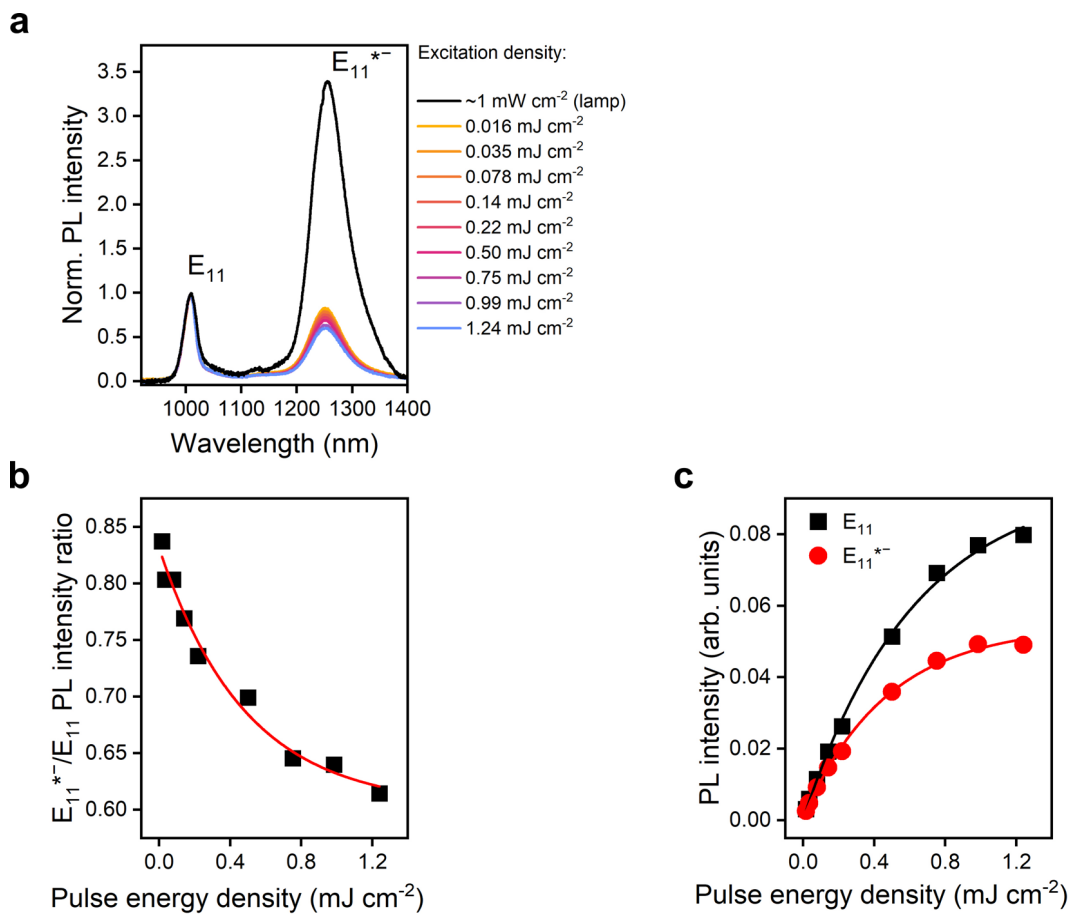

**Supplementary Figure 5. a**, Normalized photoluminescence spectra of (6,5) SWNTs functionalized with 2-iodoaniline recorded at different excitation (wavelength 575 nm) densities (pulsed) and under lamp illumination (black). **b**, $E_{11}^{*-}/E_{11}$ intensity ratios vs. pulse energy density. **c**, Absolute $E_{11}^{*-}$ (red) and $E_{11}$ (black) intensity vs. pulse energy density (lines are guides to the eye).

# Defect State Power-Dependence Depending on Optical Trap Depth

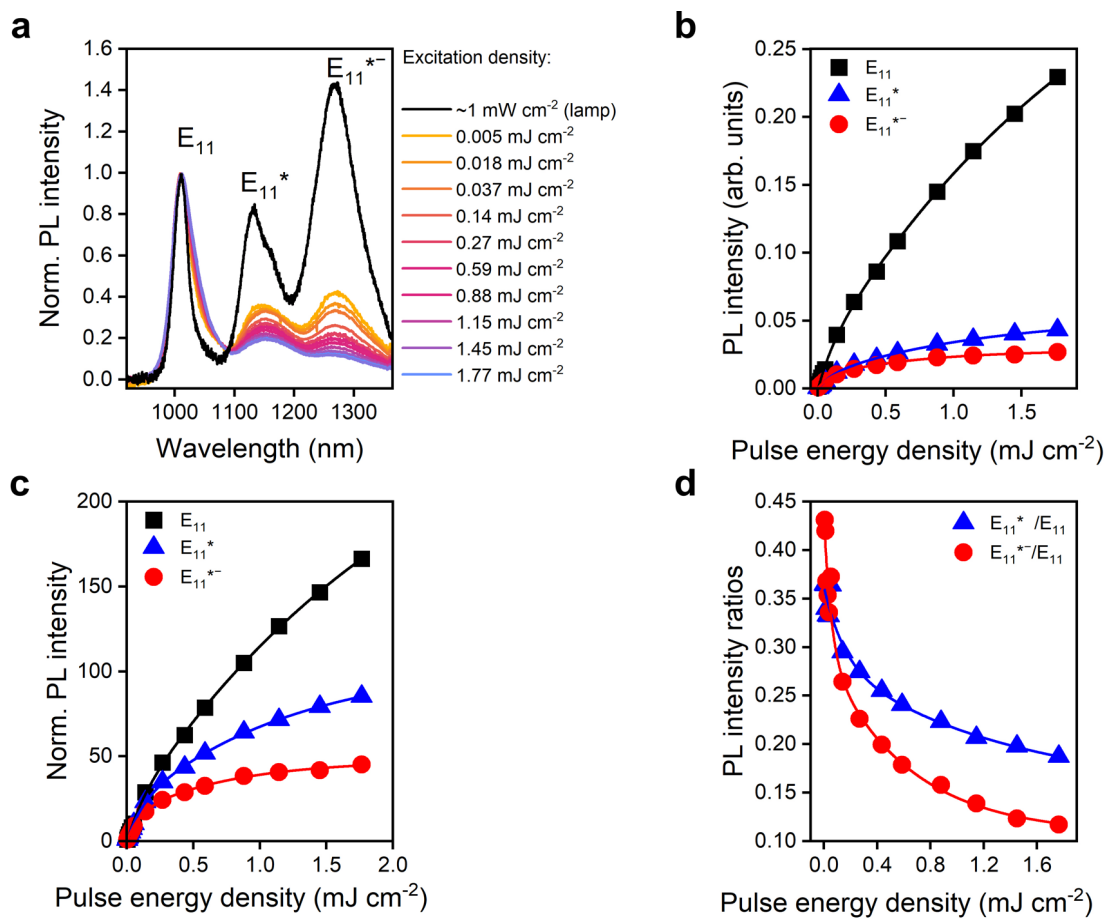

**Supplementary Figure 6. a**, Normalized PL spectra of (6,5) SWNTs functionalized with 2-iodoaniline under UV light irradiation yielding two defect emission bands with different optical trap depths labelled as $E_{11}^*$ and $E_{11}^{*-}$. Spectra were recorded at different excitation (wavelength 575 nm) densities (pulsed) and under lamp illumination. **b**, Absolute $E_{11}^{*-}$ (red), $E_{11}^*$ (blue) and $E_{11}$ (black) intensity vs. pulse energy density. **c**, Intensity of $E_{11}^{*-}$ (red), $E_{11}^*$ (blue) and $E_{11}$ (black) emission vs. pulse energy density normalized to intensity at lowest laser power. **d**, $E_{11}^{*-}/E_{11}$ (red) and $E_{11}^*/E_{11}$ (blue) intensity ratios vs. pulse energy density.

# Defect State Photoluminescence Decay Time Traces

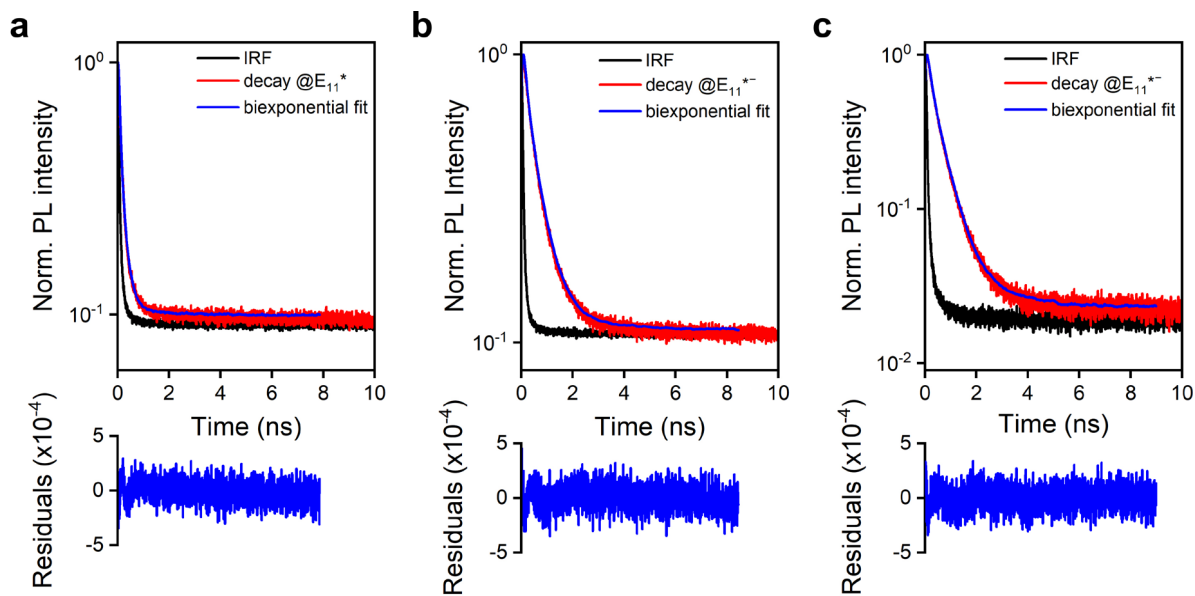

**Supplementary Figure 7. a-b**, TCSPC histograms of the photoluminescence decay (red) of (6,5) SWNT functionalized with 2-iodoaniline under UV light irradiation measured at the $E_{11}*$ (**a**) and $E_{11}*^-$ (**b**) emission maximum. **c**, TCSPC histogram of the photoluminescence decay (red) of (6,5) SWNT functionalized with 2-iodoaniline in the dark (only $E_{11}*^-$). Note that photoluminescence decays of **a** and **b** were measured on the same sample. In all cases the time traces were fitted as a biexponential decay (blue) with a reconvolution method using the fast $E_{11}$ decay as the instrument response function (IRF, black). Residuals of the fits are shown below.

# Overview of Different Defect State Photoluminescence Lifetimes

**Supplementary Table 2.** Extracted short and long lifetime components ($\tau_{short}$, $\tau_{long}$) and corresponding normalized amplitudes ($A_{short}$, $A_{long}$) for defect emission from (6,5) SWNTs functionalized with various reagents depending on the optical trap depth (energy offset between $E_{11}$ and defect emission). Corresponding defect emission bands are indicated as $E_{11}*$ and $E_{11}*^-$.

| Reagent | Optical trap depth (meV) | $\tau_{short}$ (ps) | $\tau_{long}$ (ps) | $A_{short}$ (%) | $A_{long}$ (%) |
|---|---|---|---|---|---|
| 2-iodoaniline | 247 ($E_{11}*^-$) | 134 | 568 | 54.0 | 46.0 |
| 2-iodoaniline | 136 ($E_{11}*$) | 33 | 179 | 86.7 | 13.3 |
| 2-bromoaniline | 241 ($E_{11}*^-$) | 127 | 589 | 45.5 | 54.5 |
| 5-fluoro-2-iodoaniline | 244 ($E_{11}*^-$) | 108 | 555 | 55.4 | 44.6 |

# Supplementary Note 2: Temperature Dependence of Defect State Photoluminescence

To investigate the impact of temperature on defect emission we recorded PL spectra of thin films of (6,5) SWNTs functionalized with 2-iodoaniline (in the dark and under UV illumination) from 4 K to 330 K (Supplementary Figures 8 and 9) in an optical cryostat as well as of dispersions between 278 K and 308 K (Supplementary Figures 10 and 11) in a Peltier-based temperature-controlled cuvette holder (Fluorolog). Note that the relative intensity of defect emission varies strongly between PL measurements of dispersions and thin films (see Supplementary Figures 9 and 11). This effect can be assigned to the power-dependence of the defect state emission compared to $E_{11}$, which is more pronounced for $E_{11}^{*-}$ defects (see Supplementary Figure 6).

Starting at 4 K, the nanotube thin films showed a relative increase in defect emission with increasing temperature (both for $E_{11}^*$ and $E_{11}^{*-}$) reaching a maximum between 180 K and 220 K. This increase in defect emission correlates well with the model of a potential barrier around the defect site as proposed by Kim et al.[4] In this context, the increase in defect emission is attributed to a higher exciton trapping efficiency. After reaching this maximum, the defect emission steadily decreased again. This decrease in defect emission was investigated in more detail for dispersions and closer to room temperature.

The temperature-dependent distribution of localized and mobile excitons is commonly associated with a certain detrapping energy of the localized excitons, which is however quite different from the optical trap depth. As described by Kim et al. the thermal detrapping energy $\Delta E_{thermal}$ can be determined from a van't Hoff plot of the ratio of the integrated PL intensities ($I_{E_{11}}$ and $I_{E_{11}^*}$ or $I_{E_{11}}$ and $I_{E_{11}^{*-}}$) at different temperatures ($T$):[5]

$$\frac{I_{E_{11}}}{I_{E_{11}^*}} \propto e^{-\left(\frac{\Delta E_{thermal}}{kT}\right)} \tag{1}$$

$$\ln\left(\frac{I_{E_{11}}}{I_{E_{11}^*}}\right) = -\frac{\Delta E_{thermal}}{kT} + A \tag{2}$$

where $k$ is the Boltzmann constant and $A$ is a correction factor. The van't Hoff plots for the $E_{11}^*$ and $E_{11}^{*-}$ emission bands display good linear fits to the data and extraction of $\Delta E_{thermal}$. According to this analysis the $E_{11}^*$ defect exhibits a detrapping energy of approximately 79 meV, while the apparent detrapping energy for $E_{11}{*}^-$ is surprisingly low with ~25 meV (see Supplementary Figures 10 and 11). Note that for the determination of the detrapping energy of $E_{11}{*}^-$ defects, samples with different defect densities were used. Kim *et al.* previously reported that the detrapping energy increases at higher defect density.[5] While a similar trend can be observed here, the detrapping energy of $E_{11}{*}^-$ defects increases only slightly from 23 to 27 meV. Such a low detrapping energy is in clear contrast to the increased defect state PL lifetime of $E_{11}{*}^-$. Previously, the strong correlation between defect state lifetime and optical trap depth was suggested to originate from phonon-assisted thermal detrapping.[6] Thus, we expect that a different temperature-dependent non-radiative decay mechanism, such as multiphonon decay, is dominant for $E_{11}{*}^-$ defects within the high temperature range. This finding further highlights the complex relationship between optical trap depths and thermal detrapping.

# Temperature Dependent Defect State PL in Thin Films

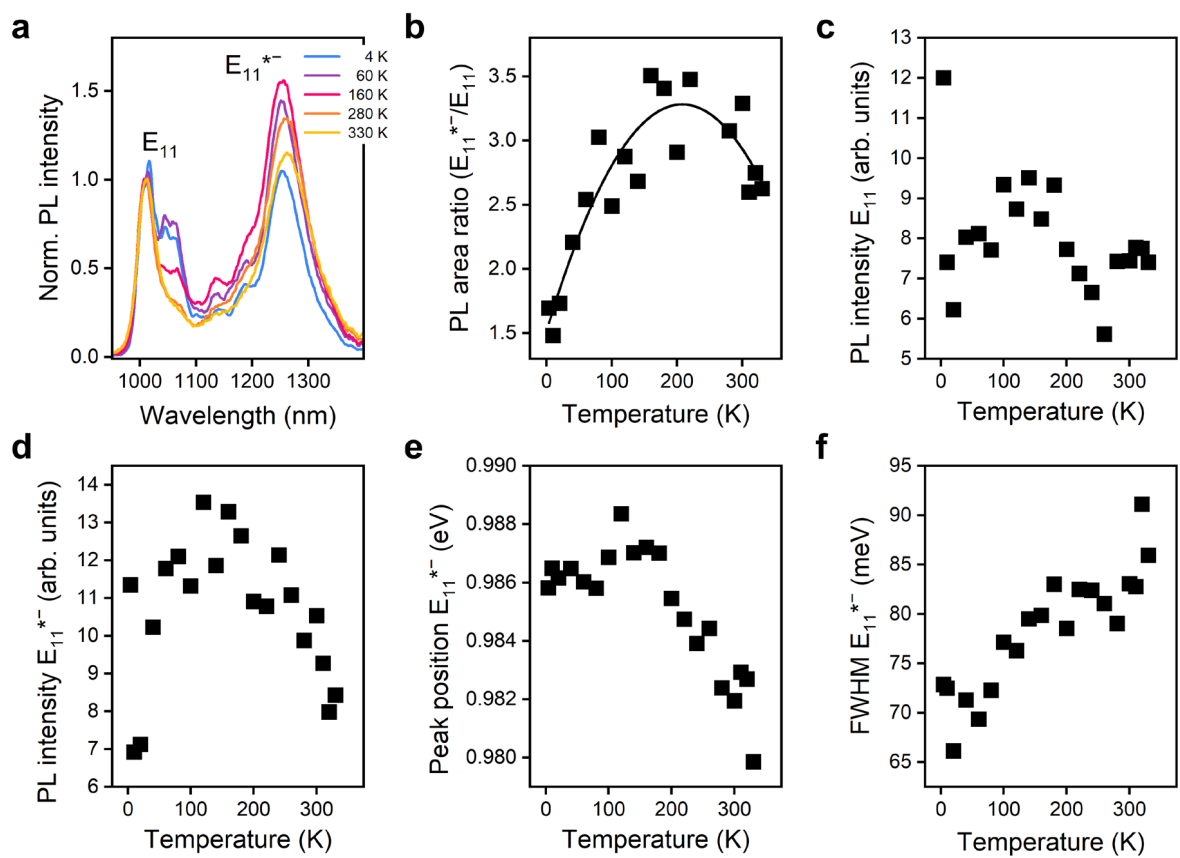

**Supplementary Figure 8. a**, Temperature-dependent (4 K – 330 K) defect state photoluminescence of thin films of (6,5) SWNTs functionalized with 2-iodoaniline in the dark. **b**, Integrated $E_{11}*^-/E_{11}$ emission ratios vs. temperature. The defect emission ($E_{11}*^-$) reaches a maximum around 200 K relative to the $E_{11}$ emission. Solid line are guides to the eye. **c-f**, Temperature-dependent change of absolute $E_{11}$ emission (**c**), absolute $E_{11}*^-$ emission (**d**), peak position of $E_{11}*^-$ emission (**e**) and full width at half maximum (FWHM) of $E_{11}*^-$ emission peaks (**f**).

Note, emission features around 1050 nm are not exclusively observed for functionalized SWNT but also for pristine SWNTs as reported by Kadria-Vili *et al.* and commonly labelled as $Y_1$ band.[7] It is assumed that this feature originates from defects that are unintentionally introduced during processing of SWNTs. The observed intensity can therefore vary even for ("pristine") SWNTs before functionalization. The functionalization of SWNTs represents an additional processing step with varying reaction times, sonication and heating steps (for SWNT films as presented in Supplementary Figure 8a) and thus can lead to even stronger batch-to-batch variations. The temperature dependence of these features is not yet understood.

**Effect of Optical Trap Depth on Temperature Dependent Defect PL (Film)**

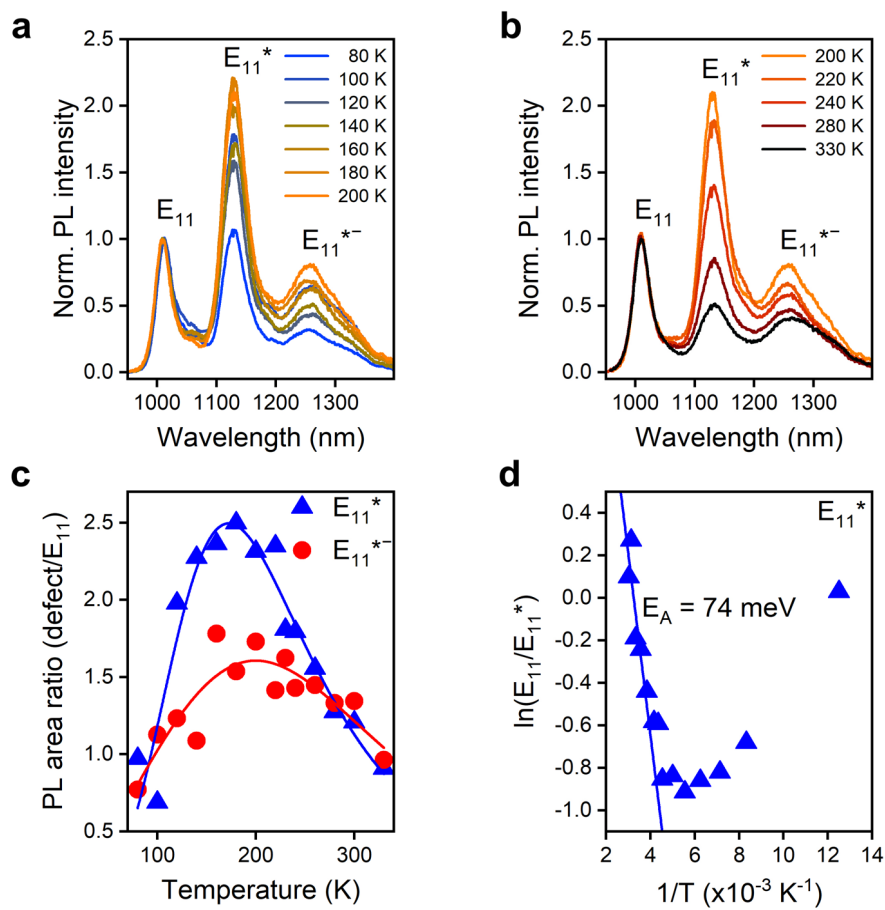

**Supplementary Figure 9. a-b**, Temperature dependent defect state PL of films of (6,5) SWNTs functionalized with 2-iodoaniline under UV light irradiation. Defect state PL is increasing relative to $E_{11}$ emission up to 200 K (**a**) and decreasing from 200 K to 330 K (**b**). **c**, Integrated $E_{11}^*/E_{11}$ (blue) and $E_{11}^{*-}/E_{11}$ (red) emission ratios vs. temperature. Solid lines are guides to the eye. **d**, van't Hoff plot for the $E_{11}^*$ emission band and linear fit to the data. A detrapping energy ($\Delta E_{thermal}$) of 74 meV was determined.

**Temperature Dependent Defect State PL in SWNT Dispersion**

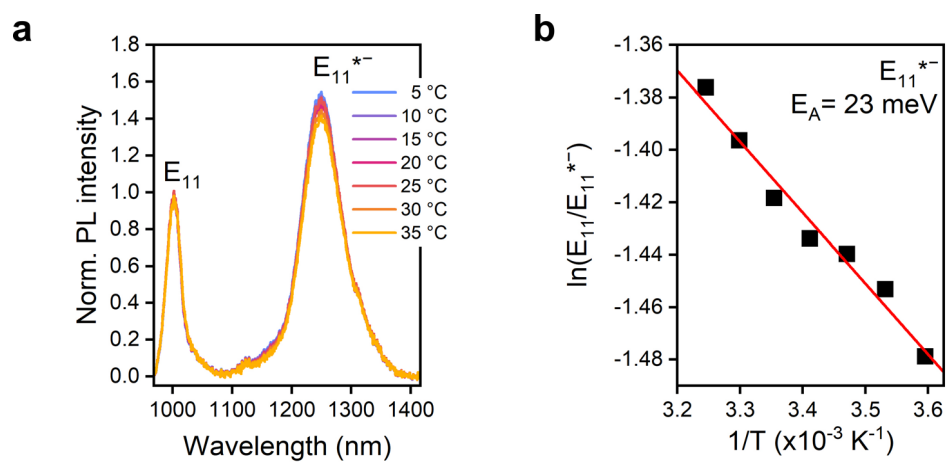

**Supplementary Figure 10. a**, Temperature-dependent PL spectra of dispersions (excitation with Xenon lamp) of (6,5) SWNTs functionalized with 2-iodoaniline in the dark. **b**, van't Hoff plot for the $E_{11}^{*-}$ emission band and linear fit to the data. A detrapping energy ($\Delta E_{thermal}$) of 23 meV was extracted. Note that the defect concentration of the functionalized (6,5) SWNTs in this measurement was lower compared to Supplementary Figure 8 (thin film) to ensure dispersion stability.

**Effect of Optical Trap Depth on Temperature Dependent Defect PL (Dispersion)**

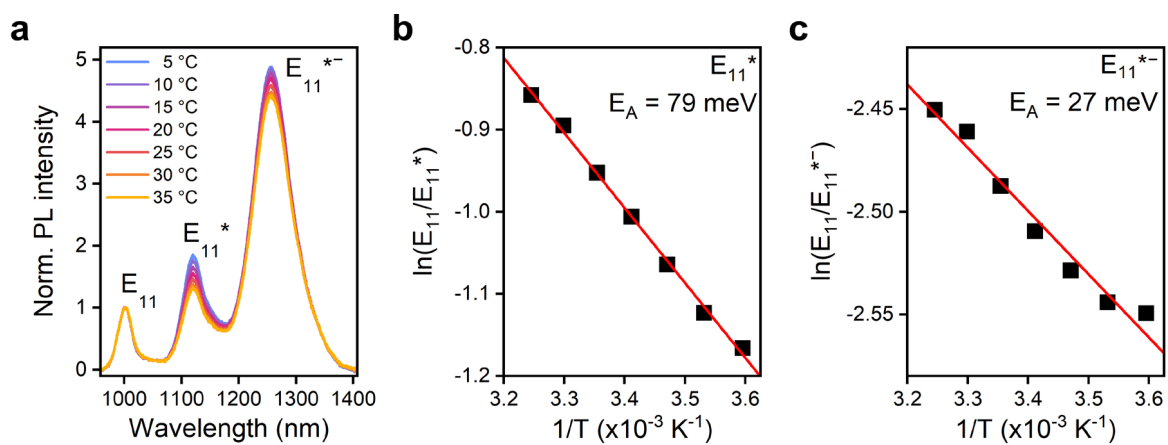

**Supplementary Figure 11**. **a**, Temperature-dependent photoluminescence spectra (excitation with Xenon lamp) of a dispersion of (6,5) SWNTs functionalized with 2-iodoaniline by UV light irradiation. **b**, van't Hoff plot for the $E_{11}^{*}$ emission band and linear fit to the data. **c**, van't Hoff plot for the $E_{11}^{*-}$ emission band and linear fit to the data. Detrapping energies ($\Delta E_{thermal}$) of 79 meV (**b**) and 27 meV (**c**) were extracted. Note that the same functionalized SWNTs were used to prepare the film in Supplementary Figure 9.

**Cumulative Photoluminescence Spectrum from Individual (6,5) SWNTs**

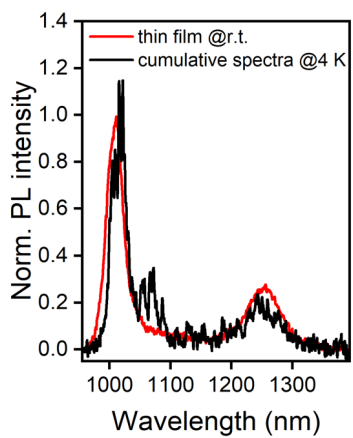

**Supplementary Figure 12.** Cumulative photoluminescence spectrum of 62 measured individual nanotubes (spots) at 4 K (black) representing the average signal of a large number of functionalized (6,5) SWNTs. The cumulative spectrum (black line) fits the spectrum measured for a thin film of the same functionalized (6,5) SWNTs at room temperature (r.t., red line). The higher signal at 1060 nm at low temperature is assumed to originate from defects unintentionally introduced during initial processing of SWNTs (dispersion etc.). Their number and emission intensity vary even for untreated ("pristine") SWNTs.[7]

**Absorption Characteristics of Polymer Wrapped Large Diameter SWNTs**

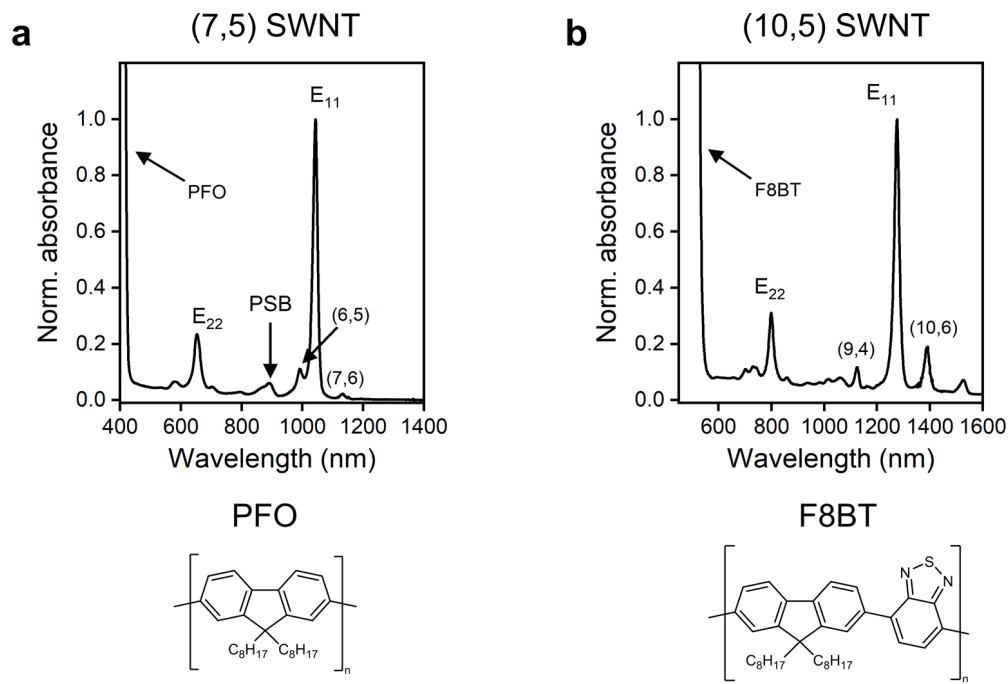

**Supplementary Figure 13. a**, Normalized absorption spectra of (7,5) SWNTs selectively dispersed in toluene with PFO. **b**, Normalized absorption spectra of (10,5) SWNTs selectively dispersed in toluene with F8BT. Characteristic nanotube transitions are indicated as $E_{11}$, $E_{22}$ and PSB (phonon sideband). Additional SWNT species and absorption bands of the wrapping polymers are indicated.

# Characterization of (7,5) SWNTs Functionalized with 2-Iodoaniline

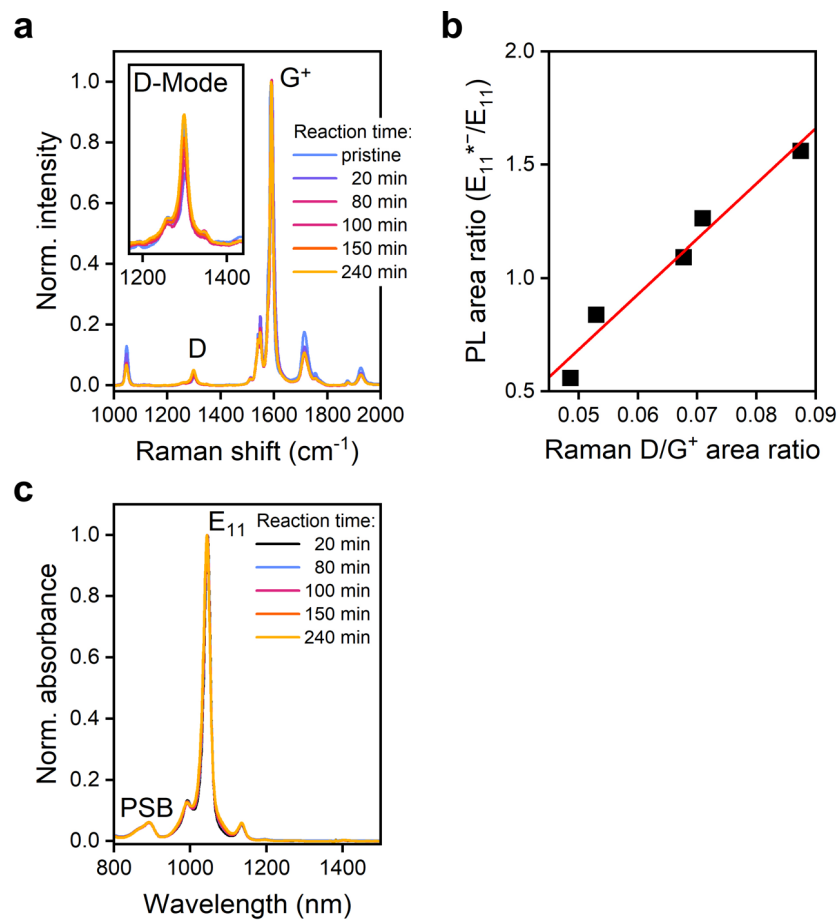

**Supplementary Figure 14. a**, Averaged Raman spectra of (7,5) SWNTs functionalized with 2-iodoaniline and after different reaction times. Inset: zoom-in on the D-mode region. **b**, Integrated $E_{11}^{*-}/E_{11}$ emission area ratios vs. integrated Raman $D/G^+$ ratios as a metric for defect density. **c**, Absorption spectra of (7,5) SWNTs functionalized with 2-iodoaniline after different reaction times. The defect density is still too low to observe significant defect absorption.

**Diameter-Dependent Defect Properties of Functionalized SWNTs**

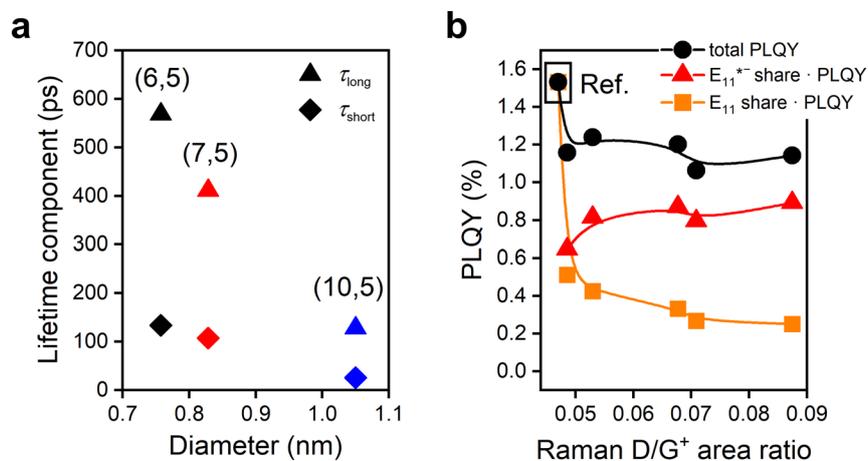

**Supplementary Figure 15. a**, Photoluminescence lifetimes (short and long component) of $E_{11}{*}^{-}$ defects of polymer-wrapped (6,5), (7,5) and (10,5) SWNTs decreasing with increasing SWNT diameter. **b**, Photoluminescence quantum yield (PLQY) of (7,5) SWNTs functionalized with 2-iodoaniline (total and spectral contributions) vs. integrated Raman D/G$^+$ ratios as metric for defect density.

# Supplementary Note 3: Mechanistic Considerations

Reference and control experiments revealed that defect emission bands located at 1130-1180 nm ($E_{11}*$) and ~1250 nm ($E_{11}*^-$) appear for various reaction conditions (see Supplementary Tables 3 and 4). As the functionalization process involves multiple components, solvents and reagents, the role and impact of each shall be summarized and discussed briefly here. It should also be noted that preference for specific defect configurations by introducing $sp^3$-defects in close proximity to already existing defect site may also play a minor role.[8]

**Toluene:**

Toluene was chosen as the main solvent because polymer-sorting of SWNTs is typically performed in this medium and the employed aniline derivatives show good solubility. This solvent choice facilitates the direct functionalization after the SWNT sorting process without further processing steps and ensures stable dispersions.

**THF:**

THF can be used to drastically speed up the rate of the functionalization reaction. While polymer-wrapped SWNTs remain stable in this solvent, the solubility of KO'Bu is greatly increased. An enhanced reactivity can be observed by comparing the reaction times of functionalization performed in toluene *vs*. those in THF. Hence, small fractions (8.3 vol%) of THF were added to the toluene reaction mixture in the standard protocol.

**Potassium *tert*-butoxide (KO'Bu):**

While KO'Bu represents a key reagent in many organic transition-metal-free coupling reactions[9-11] it has been applied for carbon nanotube chemistry only once to the best of our knowledge.[12] Upon combining aniline derivatives with the base KO'Bu, fast deprotonation of the amine group (or other acidic protons) is expected. The resulting anion may then proceed to react with the nanotube. Increasing the amount of base shifts the equilibrium of this deprotonation step and consequently increases the rate of functionalization.

It is important to note that while KO$^t$Bu is most commonly used as a strong organic base, it can also form radicals in the presence of appropriate electron acceptors.[11] As shown by the control reactions performed in this study, KO$^t$Bu itself is able to lead to emissive defect states in the $E_{11}*$ and $E_{11}*^-$ region without the presence of an aniline derivative. Thus, while the exact interactions with carbon nanotubes are unknown, it has the potential to follow nucleophilic as well as radical reaction paths with the nanotubes, possibly leading to $E_{11}*$ and $E_{11}*^-$ emission bands.

**Dimethylsulfoxide (DMSO) / Dimsyl anion:**

DMSO is a common co-solvent for KO$^t$Bu and is known to dramatically increase its basicity.[13] Hence, when functionalization is performed in the dark, the addition of DMSO may shift the equilibrium of deprotonated aniline derivatives resulting in increased functinalization rates and higher selectivity towards $E_{11}*^-$ defects. Furthermore, in the presence of small amounts of a base, a dimsyl anion can be formed that behaves like a typical carbanion and represents a strong organic base.[14] Thus, in the absence of aniline derivatives it can potentially initiate nucleophilic reaction paths leading to $E_{11}*^-$ defects with high selectivity.

Upon UV light excitation (~350 nm) the dimsyl anion can act as an electron donor and radical source.[15] Under UV illumination in the presence of DMSO we observed $E_{11}*$ emission bands, which we associate with such a radical reaction path.

Additionally, weak $E_{11}*$ bands were observed when the functionalization was performed without the addition of KO$^t$Bu under UV illumination. The absence of $E_{11}*^-$ defects is consistent with the proposed functionalization mechanism as no nucleophilic species can be formed.

**Aniline Derivatives:**

The addition of 2-iodoaniline or 2-fluoroaniline was found to greatly increase the functionalization rate for $E_{11}*$ and $E_{11}*^-$ defects when THF was used as a solvent. When functionalization is performed in toluene, vastly different reactivities and selectivities towards $E_{11}*$ and $E_{11}*^-$ defects were found compared to functionalization in the absence of aniline derivatives. Thus, a functionalization path via aniline intermediates is highly likely. It has to be noted that $E_{11}$* defects were observed for functionalization with 2-fluoroaniline, thus a dehalogenation step is not necessary for the introduction of $E_{11}$* defects.

**UV-Light Illumination:**

UV-light illumination was found to increase the reaction rate for the introduction of $E_{11}$* as well as $E_{11}$*⁻ defects. The latter is expected to originate partially from heating effects during the illumination process. Importantly, no reaction of the wrapping polymer with SWNTs was observed upon illumination.

**General Remarks about the Importance of the Base**

Several control and test reactions highlight the importance of the base in this reaction system. While a base that purely engages in a nucleophilic reaction pathway is desirable, the following aspects should be considered before testing alternative bases:

1. The highest selectivity towards the introduction of $sp^3$-defects showing $E_{11}$*⁻ emission was observed for anilines, which are very poor acids with a pk$_a$ of 28.7 (2-fluoroaniline in DMSO[16]), thus a strong base is needed for effective formation of the reactive aniline anion.
2. Strong bases such as alkyl lithium compounds lead to significant side-wall functionalization of SWNTs[17-19] and may not follow the desired functionalization path.
3. The employed base should not act as reducing agent as this can lead to a Billups-Birch type functionalization.[20]
4. To ensure high reproducibility it would be beneficial to perform the functionalization process under ambient conditions. Thus, bases that are highly sensitive toward oxygen and water should be avoided.

In summary, KO$^t$Bu in DMSO as co-solvent represents an excellent system as it is a strong base (pk$_a$ 32.2), a poor reducing agent and bench-stable in the dark.

## XPS of (6,5) SWNTs Functionalized with 5-Fluoro-2-iodoaniline

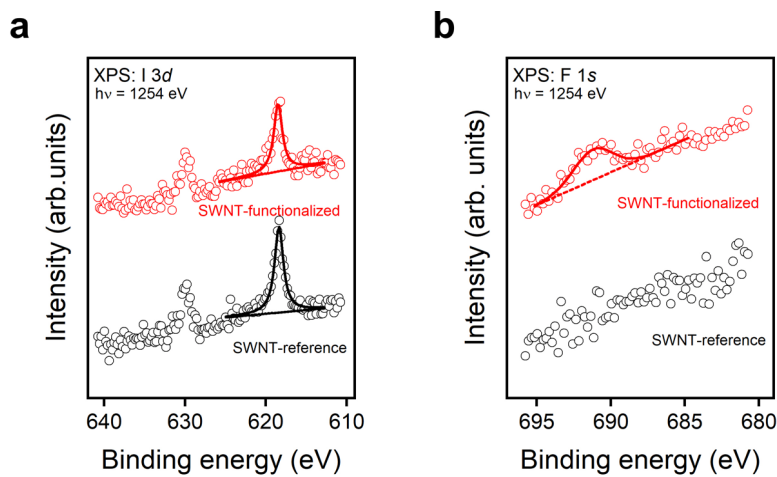

**Supplementary Figure 16. a-b**, I 3*d* **(a)** and F 1*s* **(b)** X-ray photoemission spectra of thin films of (6,5) SWNTs after reaction with 5-fluoro-2-iodoaniline with (red, high degree of functionalization) and without (black; reference) addition of KO*t*Bu as initiator. Solid lines represent peak fits. The observed binding energies of F 1*s* and I 3*d* can shift slightly due to charging effects, which cannot be excluded completely for thin films even on conducting substrates. No change in PL was observed for the reference compared to a pristine sample, indicating lack of functionalization. Note that 5-fluoro-2-iodoaniline could potentially undergo polymerization reactions in the presence of KO*t*Bu. The iodine signal for the reference may result from physisorbed iodine.

# Reference Experiments with (6,5) SWNTs in Toluene

**Supplementary Table 3.** Control reactions were performed under various conditions with (6,5) SWNTs in toluene without (red cross) and with aryl reactant (2-iodoaniline or 2-fluoroaniline), with or without UV irradiation, different KO$^t$Bu concentrations, and with or without addition of THF and DMSO to increase KO$^t$Bu solubility. The resulting photoluminescence spectra are shown and observations are summarized.

| Aryl reactant | Reaction conditions | PL Spectrum | Comments |
|---|---|---|---|
| 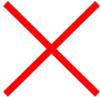 | 30 min irradiation at 365 nm | 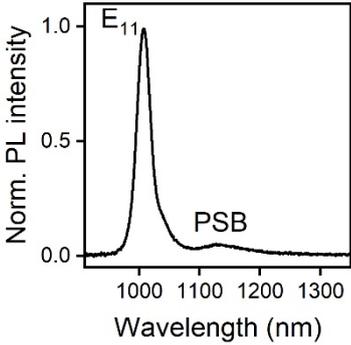 | no functionalization |
| 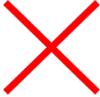 | 10 min irradiation at 365 nm, 58.60 mmol L$^{-1}$ KO$^t$Bu, 8.3 vol% THF | 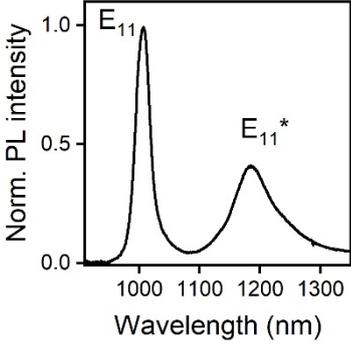 | low degree of functionalization, defect emission feature at 1185 nm |
| 2-iodoaniline 29.30 mmol L$^{-1}$ 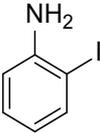 | 10 min irradiation at 365 nm, 58.60 mmol L$^{-1}$ KO$^t$Bu, 8.3 vol% THF | 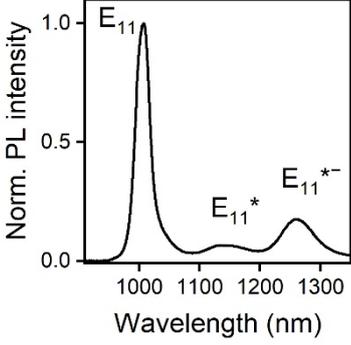 | low degree of functionalization, higher selectivity towards $E_{11}{*}^-$ emission compared to reaction without 2-iodoaniline (see above) |

| Aryl reactant | Reaction conditions | Spectrum | Comments |
|---|---|---|---|
| 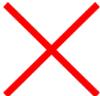 | 30 min<br>dark<br>58.60 mmol L$^{-1}$ KO$^t$Bu<br>8.3 vol% THF | 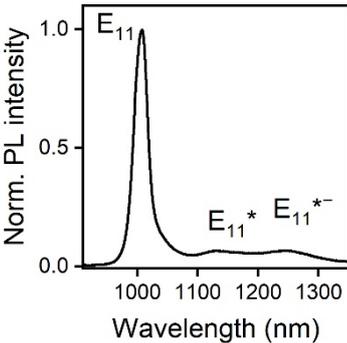 | low degree of functionalization |
| 2-iodoaniline<br>29.30 mmol L$^{-1}$<br>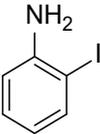 | 30 min<br>irradiation at 365 nm<br>14.65 mmol L$^{-1}$ KO$^t$Bu<br>8.3 vol% THF | 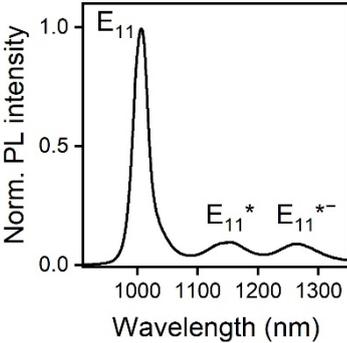 | low degree of functionalization,<br>$E_{11}^*$ can be identified as additional defect emission band |
| 2-iodoaniline<br>29.30 mmol L$^{-1}$<br>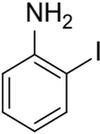 | 30 min<br>dark<br>58.60 mmol L$^{-1}$ KO$^t$Bu<br>8.3 vol% THF | 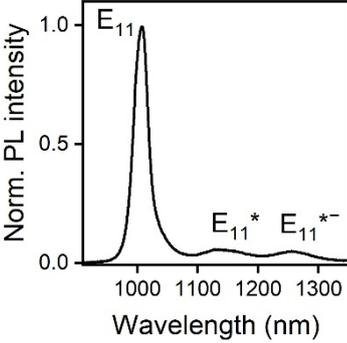 | low degree of functionalization |

| Aryl reactant | Reaction conditions | Spectrum | Comments |
|---|---|---|---|
| 2-iodoaniline<br>29.30 mmol L$^{-1}$<br>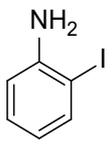 | 30 min<br>dark<br>58.60 mmol L$^{-1}$ KO$^t$Bu<br>8.3 vol% THF<br>8.3 vol% DMSO | 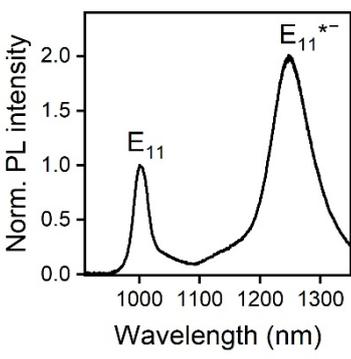 | high degree of functionalization,<br>addition of DMSO leads to a dramatic increase in reactivity,<br>high selectivity towards $E_{11}{*}^{-}$ emission |
| 2-iodoaniline<br>29.30 mmol L$^{-1}$<br>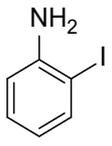 | 15 min<br>irradiation at 365 nm<br>21.98 mmol L$^{-1}$ KO$^t$Bu<br>8.3 vol% THF<br>8.3 vol% DMSO | 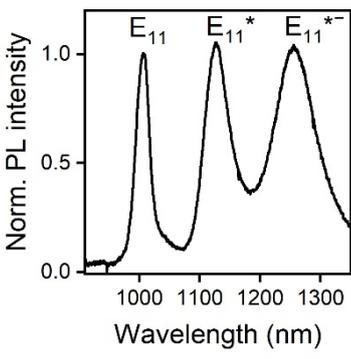 | high degree of functionalization,<br>addition of DMSO leads to a dramatic increase in reactivity,<br>both $E_{11}*$ and $E_{11}{*}^{-}$ emission band are prominent |
| 2-fluoroaniline<br>29.30 mmolL$^{-1}$<br>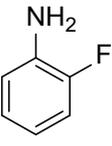 | 30 min<br>irradiation at 365 nm<br>21.98 mmol L$^{-1}$ KO$^t$Bu<br>8.3 vol% THF<br>8.3 vol% DMSO | 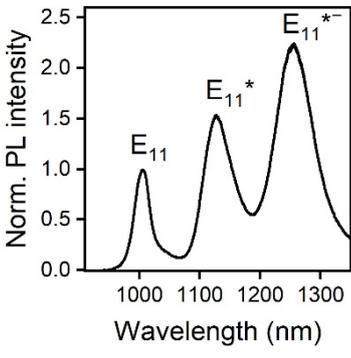 | high degree of functionalization,<br>introduction of $E_{11}*$ defect emission can also occur without dehalogenation |

# Reference Experiments with (6,5) SWNTs in THF

**Supplementary Table 4.** Control reactions were performed under various conditions with (6,5) SWNTs in THF without (red cross) and with aryl reactant (2-iodoaniline or 2-fluoroaniline), with or without UV irradiation, different KO$^t$Bu concentrations, and with or without addition of DMSO. The resulting photoluminescence spectra are shown and observations are summarized.

| Aryl reactant | Reaction conditions | Spectrum | Comments |
|---|---|---|---|
| ✗ | 10 min irradiation at 365 nm | 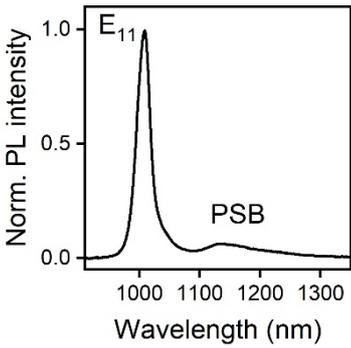 | no functionalization |
| ✗ | 10 min irradiation at 365 nm, 8.3 vol% DMSO | 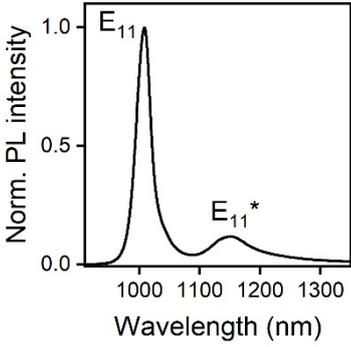 | Very low degree of functionalization, defect emission feature at 1151 nm |
| ✗ | 5 min dark, 58.60 mmol L$^{-1}$ KO$^t$Bu, 8.3 vol% DMSO | 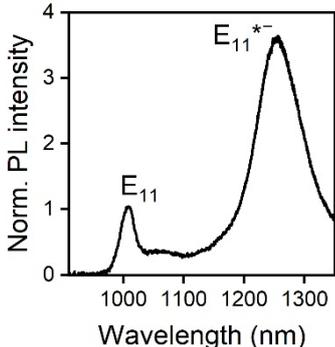 | High degree of functionalization despite absence of aniline derivative, higher reactivity compared to functionalization in toluene |

| Aryl reactant | Reaction conditions | Spectrum | Comments |
|---|---|---|---|
| ✗ | 10 min<br>irradiation at 365 nm<br>58.60 mmol L$^{-1}$ KO$^t$Bu | Norm. PL intensity vs Wavelength (nm); peaks labeled $E_{11}$, $E_{11}*$, $E_{11}*^-$ | very high degree of functionalization,<br><br>higher reactivity compared to functionalization in toluene |
| ✗ | 10 min<br>irradiation at 365 nm<br>14.65 mmol L$^{-1}$ KO$^t$Bu | Norm. PL intensity vs Wavelength (nm); peaks labeled $E_{11}$, $E_{11}*$, $E_{11}*^-$ | medium degree of functionalization,<br><br>decreasing reactivity with decreasing amount of KO$^t$Bu (see above),<br><br>large $E_{11}*$ contribution |
| 2-iodoaniline<br>29.30 mmol L$^{-1}$<br>[structure: 2-iodoaniline with NH$_2$ and I substituents on benzene ring] | 10 min<br>irradiation at 365 nm<br>14.65 mmolL$^{-1}$ KO$^t$Bu | Norm. PL intensity vs Wavelength (nm); peaks labeled $E_{11}$, $E_{11}*$, $E_{11}*^-$ | very high degree of functionalization,<br><br>higher reactivity compared to functionalization without 2-iodoaniline (see above) |

| Aryl reactant | Reaction conditions | Spectrum | Comments |
|---|---|---|---|
| 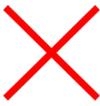 | 10 min<br>dark<br>58.60 mmol L$^{-1}$ KO$^t$Bu | 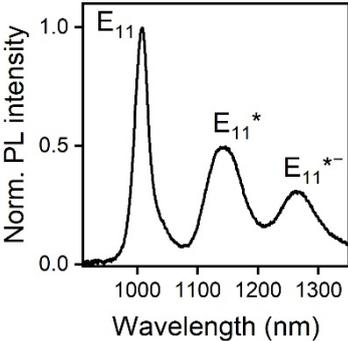 | low degree of functionalization,<br><br>introduction of $E_{11}$* defect emission bands is not only driven by UV-light |
| 2-iodoaniline<br>29.30 mmol L$^{-1}$<br>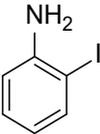 | 10 min<br>dark<br>58.60 mmol L$^{-1}$ KO$^t$Bu | 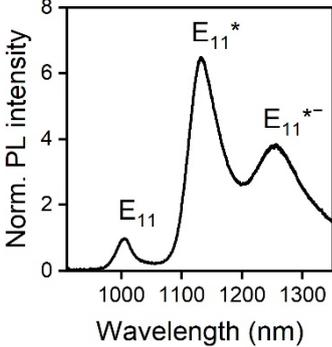 | very high degree of functionalization,<br><br>higher reactivity compared to functionalization without 2-iodoaniline (see above) |
| 2-fluoroaniline<br>29.30 mmol L$^{-1}$<br>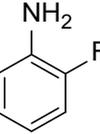 | 10 min<br>dark<br>58.60 mmol L$^{-1}$ KO$^t$Bu | 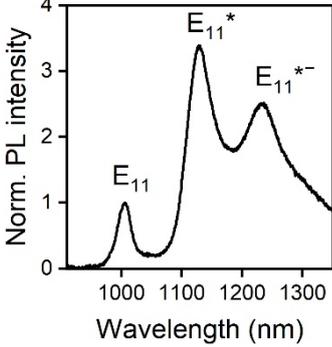 | very high degree of functionalization,<br><br>higher reactivity compared to functionalization without 2-fluoroaniline (see above) |

# Supplementary Note 4: Non-Aniline Reagents

In contrast to functionalization with aniline derivatives, a reduced reactivity and selectivity towards $E_{11}*^-$ defects is observed for non-aniline reagents. This behaviour probably originates from the molecular differences of the carbanion intermediates. The stabilization and position of the reactive carbanion can differ vastly between different substance classes. For example, Li *et al.* showed that the reaction of $C_{60}$-fullerenes with indole in the presence of KO$^t$Bu/DMSO occurs via functionalization in C3-position of indole.[21] In contrast to that, phenol attacked in the C4-position.[22] For anilines the lowest relative energy of the carbanion is expected in $C_2$-position,[23] however, steric interaction may limit an attack in this position.

Overall the lowest energy path along the potential energy surface of the reaction can vary between different reagents and lead to significant changes in reactivity for the creation of $E_{11}*^-$ defects. When the formation of $E_{11}*^-$ is reduced, other functionalization processes (e.g., radical functionalization) may be favoured kinetically. This could lead to additional shoulders and sidebands in the PL spectrum and overall lower selectivity for one specific defect emission. This concept is supported by the very similar PL spectra obtained for reactions with 2-iodophenol and thiophenol, as they represent similar substance classes und are expected to follow similar reaction paths. Lastly, for anilines the introduction of $E_{11}*^-$ defects can be controlled through the deprotonation equilibrium and thus KO$^t$Bu concentration (see Figure 1b). This equilibrium depends on the pk$_a$ values of the reagent.

**Functionalization with Non-Aniline Reagents**

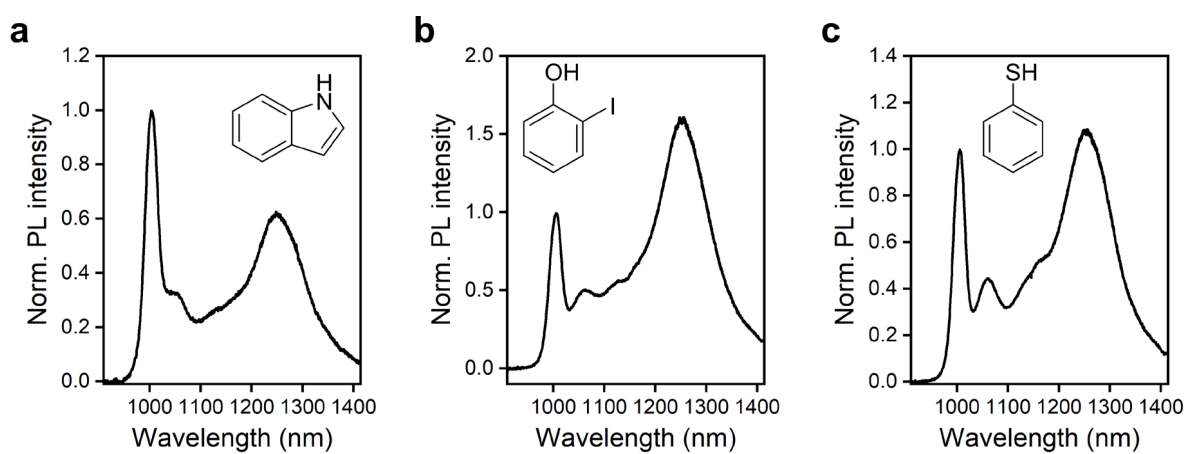

**Supplementary Figure 17. a-c**, Normalized PL spectra of (6,5) SWNTs functionalized in the dark with indole (**a**), 2-iodophenol (**b**) or thiophenol (**c**) and 2 eq. of KO$^t$Bu for 30 minutes in the dark in toluene/DMSO/THF. The concentration of indole, 2-iodophenol and thiophenol was kept at 29.30 mmol L$^{-1}$.

# $E_{11}^{*-}$ Optical Trap Depths of Functionalized (6,5) SWNTs

**Supplementary Table 5.** Summary of optical trap depths for $E_{11}^{*-}$ defects obtained by reaction of (6,5) SWNTs with seven different reagents. Different functional groups did not yield significant changes in optical trap depth.

| Reagent | Optical trap depth, $E_{11} - E_{11}^{*-}$ (meV) |
|---|---|
| 2-iodoaniline | 247 |
| 2-bromoaniline | 241 |
| 5-fluoro-2-iodoaniline | 244 |
| 2-fluoroaniline | 242 |
| indole | 244 |
| 2-iodophenol | 244 |
| thiophenol | 246 |

# Supplementary Note 5: Impact of Oxygen/Water on Functionalization

When the reaction is performed under inert conditions a strong increase in the degree of functionalization is observed. This is expected as the functionalization process may be inhibited under atmospheric conditions due to multiple effects:

(1) Quenching of the base by moisture can lead to a reduced formation of reactive carbanionic intermediates.

(2) Oxidation of negatively charged SWNT intermediates under regeneration of the carbon double bond.

(3) Oxidation of carbanionic intermediates. While the direct oxidation of the reactive carbanionic intermediate may be possible, we observed a strong dependence of the degree of functionalization on the use of DMSO as co-solvent as previously discussed. Thus, oxidation of dimsyl anions to dimethylsulfone and methanesulfonic acid is most likely.

**Functionalization under Atmospheric Conditions**

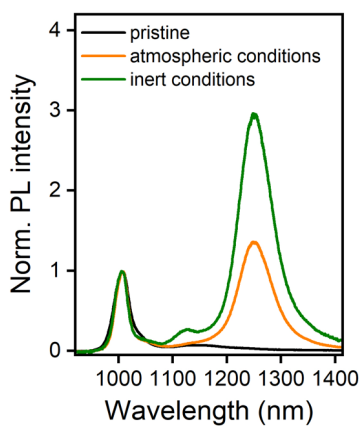

**Supplementary Figure 18.** Normalized PL spectra of (6,5) SWNTs functionalized in the dark with 2-iodoaniline/KO$^t$Bu under atmospheric conditions (orange) and under inert conditions (green). PFO-BPy wrapped (6,5) SWNTs in toluene were degassed by the freeze-pump-thaw method. A normalized PL spectrum of pristine (6,5) SWNTs is shown in black. The concentration of 2-iodoaniline was kept at 29.30 mmol L$^{-1}$.

# Supplementary References